# Electron dynamics and particle transport in capacitively coupled $Ar/O_2$ discharges driven by sawtooth up voltage waveforms

Wan Dong[1,2], Zhuo-Yao Gao[1], Li Wang[2], Ming-Jian Zhang[1], Chong-Biao Tian[1,2], Yong-Xin Liu[1], Yuan-Hong Song[1*], Julian Schulze[2]

[1]Key Laboratory of Materials Modification by Laser, Ion and Electron Beams (Ministry of Education), School of Physics, Dalian University of Technology, Dalian 116024, People's Republic of China

[2]Chair of Applied Electrodynamics and Plasma Technology, Department of Electrical Engineering and Information Science, Ruhr-University Bochum, D-44780, Bochum, Germany

[*]E-mail: songyh@dlut.edu.cn

**Abstract**

One dimensional fluid/electron Monte Carlo (MC) simulations of capacitively coupled $Ar/O_2$ discharges driven by sawtooth up voltage waveforms are performed as a function of the number of consecutive harmonics driving frequencies of 13.56 MHz, *N* (1 – 3), pressure (200 – 500 mTorr) and gas mixture (10 – 90 % admixture of $O_2$ to Ar). The effects of these external parameters on the electron dynamics, and the transport of ions and neutrals are revealed at constant peak-to-peak driving voltage. The electronegativity is found to decline as the number of consecutive harmonics increases and the DC self-bias voltage decreases. Increasing the pressure also leads to a decrease in electronegativity. The combination of a decrease in the mean free path of electrons and the presence of the Electrical Asymmetry Effect (EAE) result in different spatio-temporal distributions of the ionization rate, which lead to a reduction in the amplitude of the DC self-bias at higher pressure. As the admixture of electronegative $O_2$ increases, the electronegativity is enhanced, and the discharge mode changes from an α – Drift Ambipolar (DA) hybrid to DA mode. This work focuses on linking these fundamental changes of the plasma physics induced by changing external parameters to process relevant charged particle and neutral fluxes to the electrodes. Particular attention is paid to $O(^1D)$ flux, because it is a precursor of deposition. In discharges driven by sawtooth up voltage waveforms, placing the substrate on the grounded electrode and increasing the number of consecutive harmonics, *N*, can facilitate the deposition process, since the $O(^1D)$ flux to the substrate is higher in these scenarios. Moreover, at an $O_2$ admixture of 20%, the $O(^1D)$ flux is nearly as high as that at an $O_2$ admixture of 90%, indicating that a higher $O(^1D)$ flux can be achieved without excessively increasing the $O_2$ admixture.

## 1 Introduction

Radio frequency capacitively coupled plasmas (RF CCPs) sustained in oxygen gas or in gas mixtures containing oxygen are widely used in plasma-enhanced etching and deposition processes [1-3]. In plasma-enhanced chemical vapor deposition (PECVD) processes, ground state and excited oxygen atoms serve as precursors to deposit materials such as silicon dioxide [4,5]. In etching processes, $O_2$ also plays a crucial role in removing polymers by oxidizing them. This process reduces the loss of ions in such polymer films, allowing more ions to etch the target material and, thus, accelerating the etching process [6-9].

There are numerous studies on RF CCPs operated in oxygen and in gas mixtures containing $O_2$. These studies primarily focus on the fundamental discharge mode [10-14], charged particle heating mechanisms [15-18], or the characteristics of etching and deposition [19-21]. There are 4 well-known discharge modes, i.e., the α-, γ-, Drift-Ambipolar (DA)-, and the striation mode [10,11]. The presence of different gases as well as the choice of external control parameters are known to affect the discharge mode. In electropositive gas discharges, α- and γ-mode operation can be observed [10], whereas in mixtures of electronegative and electropositive gases, especially in highly electronegative gas mixtures, almost all discharge modes can be observed. Derzsi et al. [11] found that a significant frequency dependence of plasma parameters is present in low-pressure capacitively coupled $Ar/O_2$ discharges, which is generally considered to be indicative of transitions in the dominant discharge operating mode. Through their research, three frequency ranges can be distinguished, showing distinctly different excitation characteristics. Additionally, as the operating pressure or electrode gap are increased, the discharge mode was found to change from the DA- to the α-mode in capacitively coupled oxygen discharges [14].

A deeper understanding of the discharge mechanism is essential for knowledge based control of the discharge characteristics. To obtain such insights, a Boltzmann term analysis based on the electron momentum balance equation combined with PIC/MCC simulations was performed by Vass et al. [16]. This method confirmed previous theories and provided new discoveries. Specifically, in contrast to conventional theoretical models, which predict ‘stochastic or collisionless heating’ to be important at low pressure, they found an important contribution of ohmic power absorption. Moreover, the individual contribution of each electron power absorption mechanism was quantified numerically in a clear and concise manner, making it easier to analyze and understand the fundamental physics of the discharge.

For industrial applications, the properties and distribution functions of ions and neutrals are important, because they can directly affect the process results. In deposition processes of silicon dioxide, researchers usually consider the fluxes or densities of O atoms as well as the energies of different ion species at the electrodes in gas discharges that contain $O_2$, because O atoms are the precursors of silicon dioxide and higher ion energies can result in removal of deposited material [4,22]. For instance, Kitajima et al [21] found the fluxes of metastable $O(^1D)$ atoms to be one of the key factors to realize high quality $SiO_2$ film formation at low substrate temperatures, because this neutral particle species has the potential to enhance the oxidation process. They conducted a series of studies on $O(^1D)$ in Ar diluted $O_2$ plasmas. For etching processes, the ion energy and angular distributions, ion fluxes, and neutral fluxes are highly relevant, since these parameters directly affect the etching morphology and rate. However, there is a lack of systematic studies on establishing a connection between ions/neutral properties and discharge characteristics, and on enhanced methods to control the ion and neutral properties effectively to obtain better deposition or etching results. In our previous study [23], in capacitively coupled $Ar/O_2$ discharges driven by a single frequency, the evolution of charged and neutral particles with the change of discharge parameters was analyzed in detail based on the electron dynamics as well as the particle transport properties.

According to Moore's Law, the critical dimension of integrated circuits is shrinking year by year and there is an aggressive transition of the device architectures from two-dimensional (2D) to three-dimensional (3D), which requires that the etching and deposition processes have to be technologically advanced [24-26]. Recently, the use of electrically asymmetric driving voltage waveforms, which enable the decoupling of ion energy and ion flux control to some extent, and meanwhile expand the window of the process, is expected to be applied in etching and deposition applications to achieve flexible control of the discharge [27-32]. There are two general types of electrically asymmetric voltage waveforms, i.e., amplitude asymmetric [27-29] and slope asymmetric voltage waveforms [30, 33-34]. In capacitively coupled plasma discharges, the amplitude asymmetric voltage waveforms, such as the peaks waveform, can induce a modulation of the sheath properties near the electrode on both sides and can control the ion energy distribution at both electrodes by phase control [35-37], while the driving voltage and frequencies remain fixed, resulting in the ion flux remaining essentially unchanged [35-41]. For slope asymmetric voltage waveforms (for example, sawtooth up or down waveforms), generally, the number of harmonic frequencies is varied to control the electron kinetic properties and hence the ionization/excitation reaction rates [42-43]. In consequence, the ion fluxes are modulated while maintaining the ion energy essentially unchanged [42-47]. In addition to this, electrically asymmetric voltage waveforms have many further advantages. It is found in experiments and simulations that the radial distribution of the plasma becomes significantly more uniform when modulating the phase between the driving harmonics of the electrically asymmetric voltage waveforms [28]. There is even a strong reversed electric field near the sheath collapse phase induced by adjusting the electrically asymmetric voltage waveform, which could accelerate electrons into high aspect ratio dielectric etching trenches to eliminate wall charging effects [29]. The use of electrically asymmetric voltage waveforms in discharges can generally modulate the characteristics of the electron dynamics, which further affects the collisional reaction rates involving electrons [48,49]. Such EEDF control might provide the opportunity to selectively control the density and fluxes of charged particles and neutrals relevant to etching and deposition by Voltage Waveform Tailoring (VWT), leading to improved etching and deposition results.

Therefore, in this study, electrically asymmetric driving voltage waveforms synthesized from different numbers of consecutive harmonics of 13.56 MHz (sawtooth up waveforms) are used and their effects on the discharge characteristics as well as on the transport of charged and neutral particles are revealed at different pressures and gas mixture ratios in Ar/$O_2$ CCPs using a one-dimensional fluid/electron Monte Carlo (MC) hybrid model. The structure of the article is as follows: Section 2 provides a brief introduction of the simulation model, Section 3 delves into the results' analysis, and concluding remarks are presented in the final section.

## 2 Model description and validation

Discharges driven by sawtooth up voltage waveforms synthesized from multiple consecutive harmonics of 13.56 MHz and operated in Ar/$O_2$ gas mixtures are investigated as a function of the number of harmonics, pressure, and $O_2$ admixture using a one-dimensional fluid simulation coupled with an electron MC model. During the first five cycles of the fundamental frequency, only the fluid model is operated. The fluid model used in this stage contains the continuity equations for electrons and ions, the electron drift-diffusion approximation, the ion momentum balance equation, the neutral diffusion equations, the electron energy balance equation, and Poisson's equation. Detailed descriptions of these equations are available in our previous works [23, 50-54]. These five initial cycles of the fluid model yield an initial spatio-temporally resolved electric field, which is then used to drive the subsequent operation of the electron MC model. After these first five fundamental frequency cycles, the space-time evolution of the electric field within the fifth cycle is stored and transferred to the electron MC model. Superparticles are then randomly scattered over the computational region. In the electron MC model, the motion of these superparticles is tracked within the electric field and electron-neutral collisions are considered. Simultaneously, data on the velocity, position, and other characteristics of the electrons are recorded at each time step. This process is repeated five or more times to ensure that enough particles are tracked to minimize numerical errors. In this way the space and time resolved electron energy distribution function (EEDF) is statistically obtained. Using the EEDF and other information such as collision reaction cross sections, parameters such as the electron temperature, reaction rate coefficients, and electron transport coefficients can be determined. These coefficients and the electron temperature are then fed back into the fluid model. In the next stage, the electron energy balance equation of the fluid model is no longer used. Instead, the electron temperature calculated from the electron MC model is utilized. After the fluid model completes its operation, the updated electric field is transferred back to the electron MC model. This creates a loop, which is continuously operated for approximately 40,000 fundamental frequency cycles until the densities of charged and neutral particles, as well as all other physical parameters converge. Eventually, the code stops running, and the data are available for analysis.

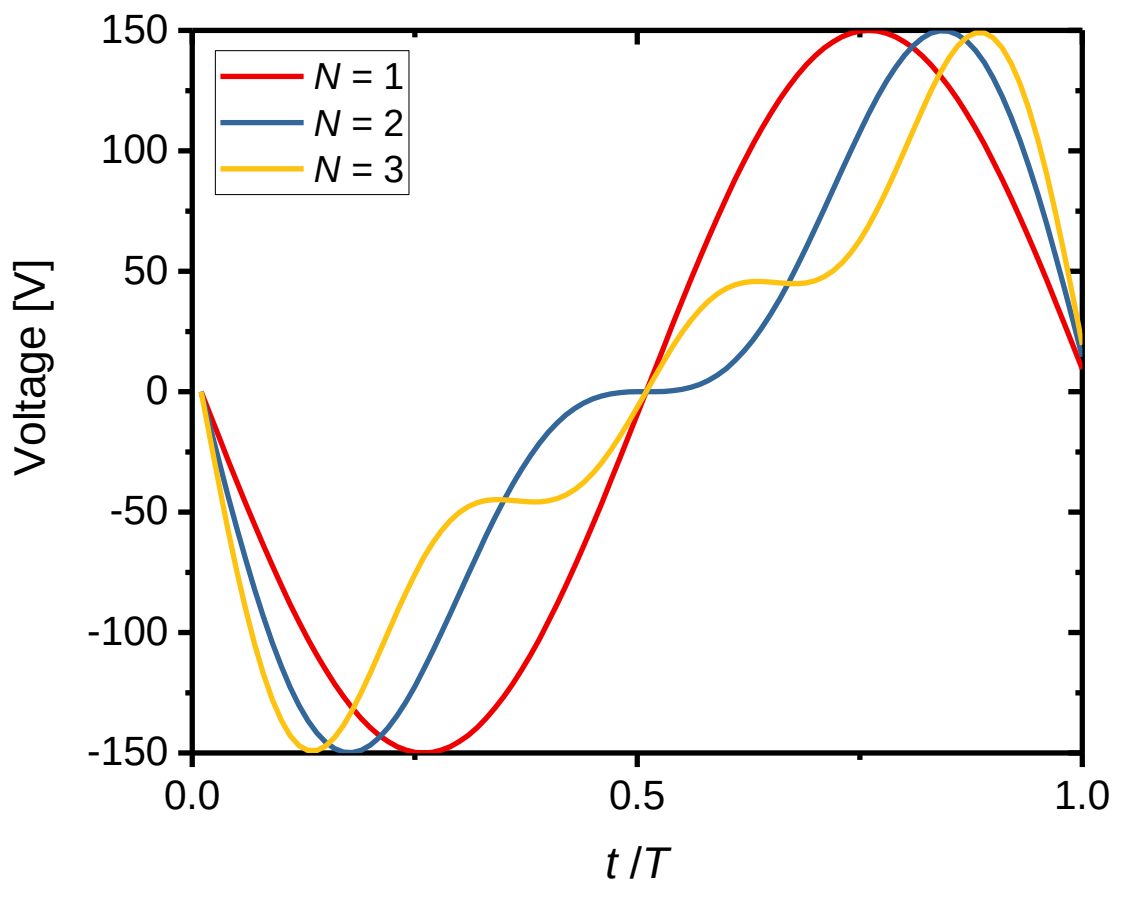

**Figure 1** Sawtooth up voltage waveforms synthesized from different numbers of consecutive harmonics, *N*, of a fundamental frequency at a fixed peak-to-peak voltage of 300 V.

Furthermore, the electric potentials at the powered electrode ($x$ = 0) and the grounded electrode ($x$ = d) satisfy $V(t)|_{x=0} = -V_0 \sum_{k=1}^{N} \frac{1}{k} \sin(k\omega t)$ and $V(t)|_{x=L} = 0$, respectively, where the voltage amplitude $V_0$ is 150 V, the number of harmonic frequencies $N$ is 1, 2, 3, and the fundamental frequency is 13.56 MHz. The voltage waveforms are shown in figure 1. Secondary electron effects are not considered in this work, since they are known not to play an important role under the voltage and pressure conditions used in this work [55, 56]. The reaction set used to describe the $Ar/O_2$ gas mixture includes a total of 56 collision reactions, encompassing elastic collisions, ionization collisions, excitation collisions, and dissociative attachment collisions. In addition to electron collisions with neutrals, the model also considers ion-ion, ion-neutral, and neutral-neutral collisions. The charged particles included in the model are electron ($e^-$), $Ar^+$, $O_2^+$, $O^+$, $O^-$. Moreover, there are four neutrals ($O(^3P)$, $O(^1D)$, $O_2(a^1\Delta_g)$, $Ar^*$) considered in the model. Since $O(^1D)$ is a precursor to the deposition of materials such as $SiO_2$ [4, 22], the transport of $O(^1D)$ will be discussed in detail. And the transport of $O(^3P)$ and $Ar^*$ is also analyzed. In this work, the diffusion coefficient of neutrals $D_n$ is considered based on the Blanc law [50], which is determined by the corresponding coefficients in the background gas Ar and $O_2$. The detailed expression of $D_n$ and relevant parameters can be found in references [51, 52], respectively. The electron reflection probability is fixed at 0.25 [53]. For detailed information on the hybrid model, the reaction set, and relevant coefficients such as the sticking coefficient of neutrals, please refer to our previous work [23, 50-54]. Additionally, it should be mentioned that due to the different time scales between the neutrals and electrons/ions, a larger time step for neutrals is used to reduce the computation time. The following convergence criterion is fulfilled for all types of particles in our simulations, respectively: $(N_{T+1}- N_T)^2/N_0^2 < 10^{-6}$, where $N_T$ is the time averaged density in the previous fundamental cycle, $N_{T+1}$ is the time averaged density in the next fundamental cycle, and $N_0$ is a typical particle density achieved after convergence. Furthermore, the hybrid model used in this work has been validated by experiments performed in the reactor described in detail in [55,56] in $Ar/O_2$ gas mixtures. Figure 2 shows simulation and experimental results for the spatio-temporally resolved excitation/ionization dynamics under the following discharge conditions: single frequency driving voltage with a frequency of 13.56 MHz and a voltage amplitude of 150 V; the electrode gap is 2.5 cm; the gas pressure is 100 mTorr with an Ar admixture of 70%, 50%, and 30%, respectively. Since the electron energy threshold (13.5 eV) as well as the shape of the energy dependent cross section for electron impact excitation from the Ar ground state to the Ar 2p1 state, which results in plasma emission at 750.4 nm are similar to those of the Ar ionization ($Ar+e \rightarrow Ar^+ +2e$) in $Ar/O_2$ (Threshold: 15.6 eV), the spatio-temporal ionization rate from the simulations is compared to the Ar excitation rate obtained from experiments.

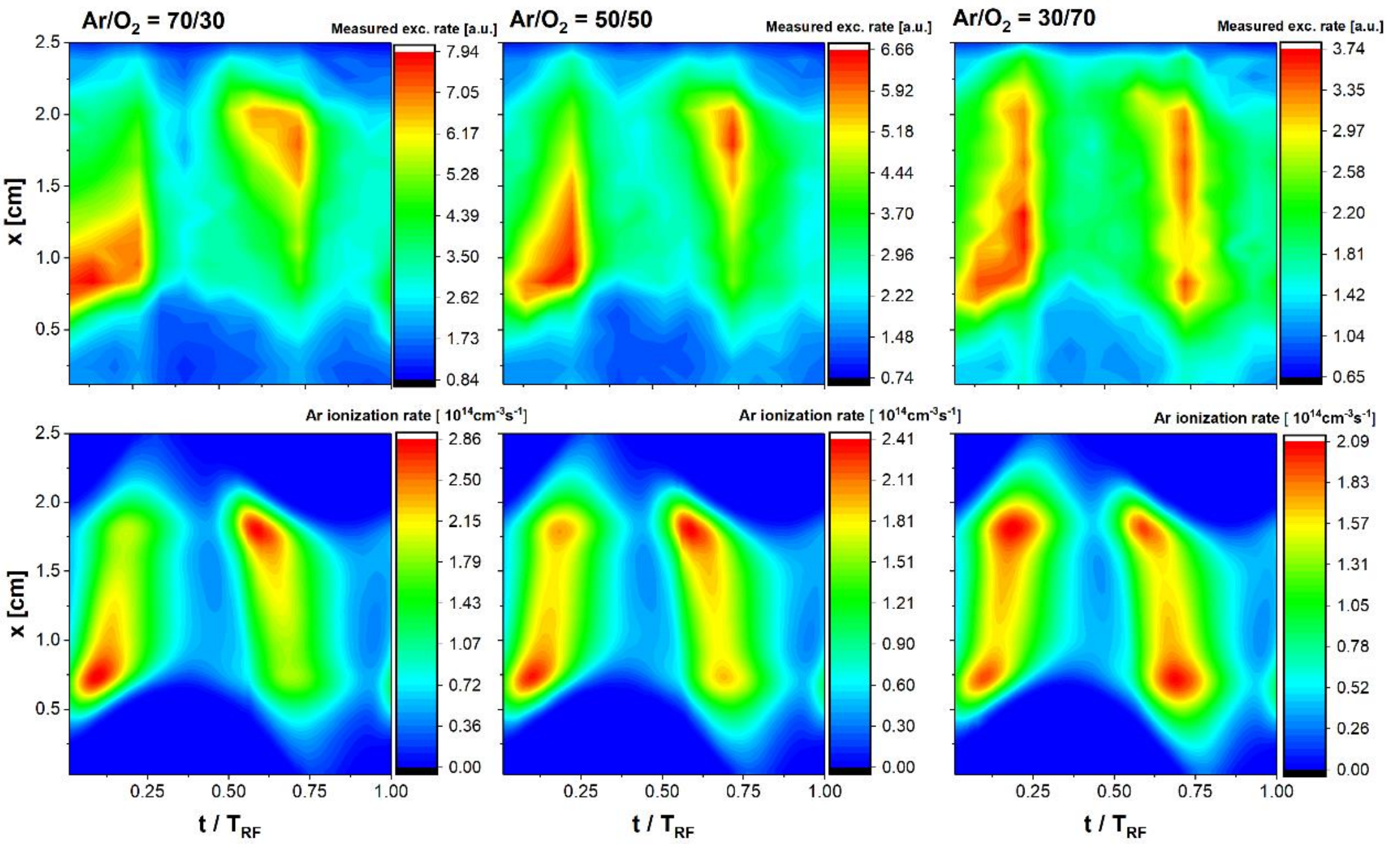

**Figure 2** First row: experimental results of the electron impact excitation rate from the ground state into the Ar2p1 state; Second row: simulation results of the total ionization rate at different $Ar/O_2$ gas ratios. Discharge conditions: $Ar/O_2$ (70/30, 50/50,30/70) mixed gases at 100 mTorr with an electrode gap of 2.5 cm, driven by a single frequency voltage with a frequency of 13.56 MHz and a 150 V voltage amplitude.

As shown, at a low $O_2$ content, i.e. $Ar/(Ar+O_2)$ = 70%, most of the excitation occurs during the sheath expansion in the experiments. With increasing the $O_2$ content, more excitation occurs in the bulk region and at the collapsing sheath edges, which indicates an α- mode to α- and DA- hybrid mode transition. This transition is properly captured by our simulations, as shown in the second row of figure 2. However, due to the slight geometrical asymmetry of the chamber in the experiment, a DC self-bias voltage is generated. This leads to a higher excitation rate during the sheath expansion at the bottom electrode ($x$ = 0 cm) compared to that at the top electrode ($x$ = 2.5 cm), which is not happening in the simulation results due to the geometrical symmetry assumption. Achieving a complete symmetry in the experiments is challenging. These results validate the reliability of the model used in this work.

## 3 Results and discussions

In this section, the discharge characteristics and transport of particles will be studied in a capacitive RF plasma operated in Ar/$O_2$ (90/10~10/90) at 200, 300, and 500 mTorr. The sawtooth up voltage waveforms shown in figure 1 are employed at a fundamental frequency of 13.56 MHz and a voltage amplitude of 150 V. The electrodes are parallel to each other with a gap of 3 cm.

### 3.1 Effects of the number of harmonic frequencies *N* on the discharge characteristics

The spatial and temporal evolutions of the electron density, electric field, electron power absorption rate, and total ionization rate are shown in figure 3, respectively, at a pressure of 200 mTorr, an Ar/$O_2$ gas mixture of 10/90, and for harmonic numbers $N$ of 1, 2 and 3. The total ionization rate is the sum of all reaction rates which generate electrons. The specific reaction set can be found in our previous study [23]. Figure 4 shows the time averaged spatial evolution of the EEDF at different numbers of driving harmonics, $N$. From figure 3, the spatio-temporally resolved plots of both plasma parameters exhibit characteristics of electronegative discharges, i.e. high electron densities near the sheath edge and high negative ion densities in the bulk. Figures 3 (a1)-(a3) show that the electron density increases at the sheath edge and in the bulk region as a function of $N$. Meanwhile, the peak of $O^-$ density increases slightly from $5.28\times10^9$ cm$^{-3}$ to $6.44\times10^9$ cm$^{-3}$ and $6.60\times10^9$ cm$^{-3}$. This change results in a slight decrease of the electronegativity from 11.95 to 10.96 and 9.42 as $N$ increases from 1 to 3. In addition, a shift in the electron and negative ion density distributions towards the upper electrode in discharges driven by multiple frequencies is observed as compared to the case of single frequency operation, which is caused by the Electrical Asymmetry Effect (EAE) present for $N > 1$. Based on the discharge condition used in this work, the variation of the number of harmonics is found to have very little effect on the $O_2(a^1\Delta_g)$ density.

Figure 3 (b) shows a decrease of the electric field in the bulk region and a change in the spatio-temporal distribution of the peaks of the electric field as the number of harmonic frequencies $N$ increases from 1 to 3 due to the change of the driving voltage waveform. Taking the electric field in the bulk region as an example, a positive peak of the electric field occurs at the time of about 0.6 $T$ within one fundamental driving period in the single frequency case. A strongly negative field is observed at 0.1 $T$ near the grounded electrode in the single frequency case. This field increases as a function of $N$ and spreads out over the entire bulk region. At $N$ = 3, there is a negative peak of the electric field near the powered electrode at 0.1 $T$ and a positive peak of the electric field occurs at 0.75 $T$. The difference in the distribution of the electric field is closely related to the driving voltage waveform and the differences in the distribution of the electron density as $N$ is increased. For the single frequency discharge, the low electron density and low conductivity cause an enhanced drift electric field in the bulk region. In the dual-frequency case, a strong ambipolar field is generated at 0.1 $T$ near the grounded electrode due to the enhanced local electron density gradient in the electropositive edge region (see figure 3 (a2)). For $N$ = 3 the electron density in the bulk region increases significantly. Thus, the bulk drift electric field is low. The rapid expansion of the sheath due to the rapid drop of the driving voltage at 0.1 $T$ causes the electric field to be high at the same time near the powered electrode. This happens, because the sawtooth-up driving voltage waveform leads to a strong electron current at this time within the fundamental RF period, when the sheath expands quickly at the powered electrode, while the current is much lower during the remainder of the fundamental RF period, when the sheath collapses slowly at the powered electrode. Specifically, as the voltage waveform changes from a single to a triple frequency waveform, the maximum electron conduction current density near the sheath expansion phase increases from 18 $A \cdot m^{-2}$ to 42 $A \cdot m^{-2}$, while in the vicinity of the sheath collapse, the minimum electron conduction current density decreases from -18 $A \cdot m^{-2}$ to -19.6 $A \cdot m^{-2}$.

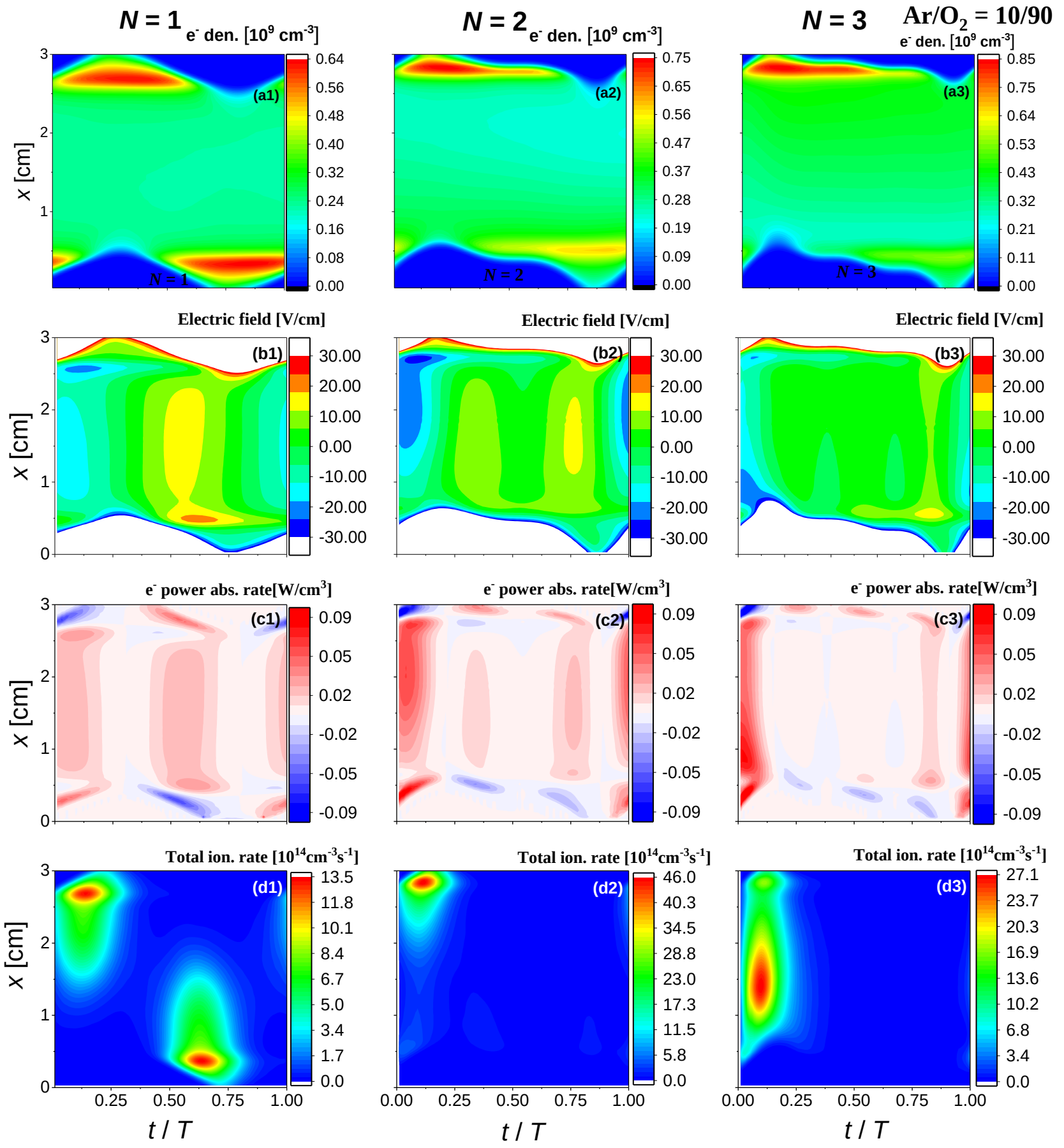

**Figure 3** Spatio-temporal evolutions of the electron density (a1)-(a3), the electric field (b1)-(b3), electron power absorption rate (c1)-(c3), and total ionization rate (d1)-(d2) as a function of the number of harmonic driving frequencies $N$ used to synthesize the sawtooth up voltage waveforms. Discharge conditions: $Ar/O_2$ (10/90) mixed gases at 200 mTorr with an electrode gap of 3 cm, sustained by single or multiple frequencies. The fundamental frequency is 13.56 MHz and the driving voltage amplitude is 150 V.

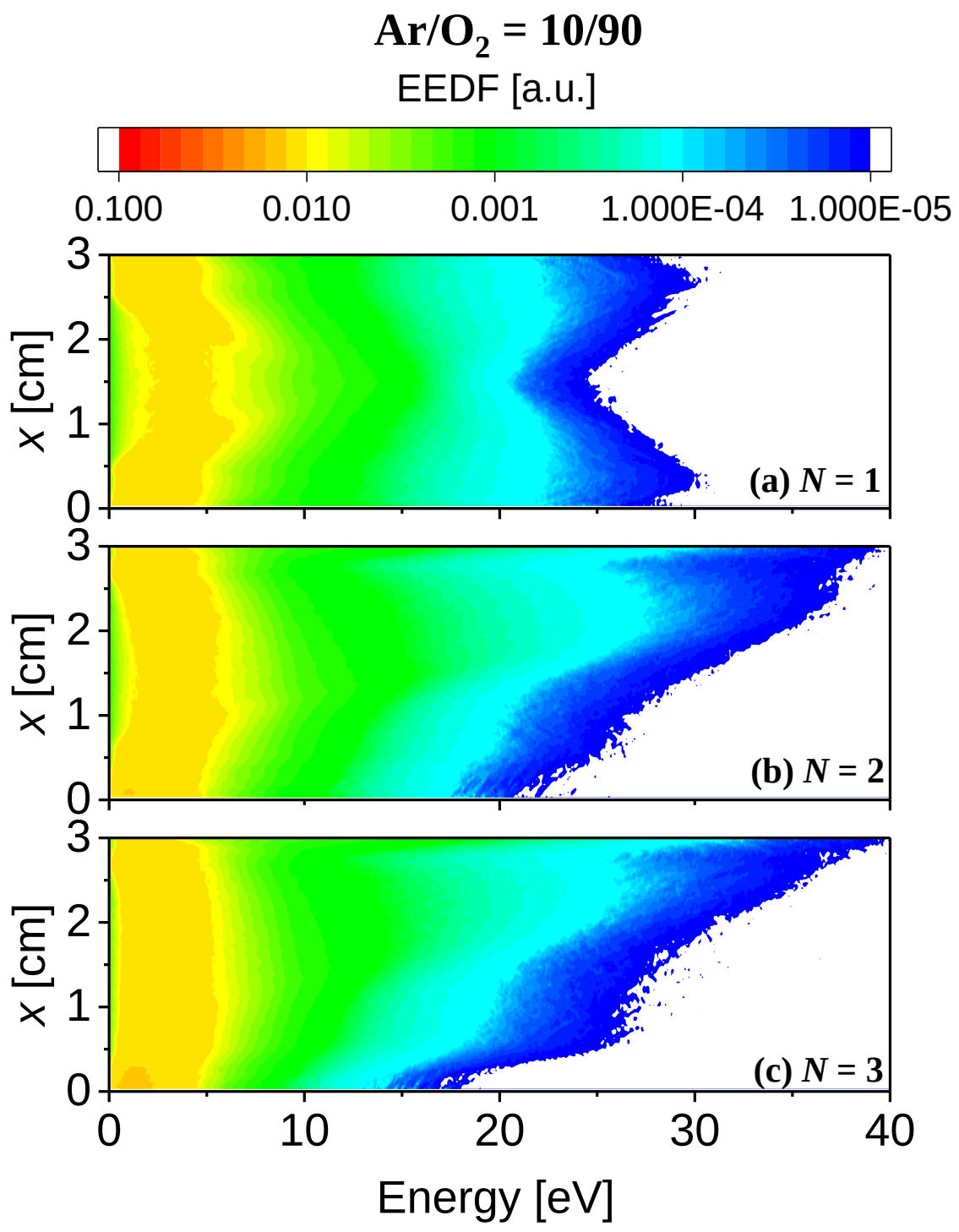

**Figure 4** Spatially resolved and time averaged EEDF for (a) $N = 1$, (b) $N = 2$, and (c) $N = 3$. The discharge conditions are the same as those in figure 3.

The electric field distribution affects the electron power absorption as well as the ionization dynamics. Figure 3 (c1) shows that, in the single frequency scenario, both the electron heating and the electric field are intense at 0.15 $T$ and 0.6 $T$. For the case of $N = 2$, the amplitude of the electron power absorption rate near the sheath expansion at the powered electrode, in the bulk region, and near the sheath collapse at the grounded electrode at 0.1 $T$ is significantly higher than that in the single frequency case, i.e., more electrons gain enough energy to ionize, and their energy is higher compared to the single frequency case. This results in a peak ionization rate at the grounded electrode during the local sheath collapse phase that is 4.7 times higher than that in the case of a single driving frequency. However, there is no strong ionization near 0.75 $T$ due to the slow expansion of the sheath near the grounded side and the limited heating of electrons by the electric field in the bulk region. But the ionization at 0.1 $T$ is high enough to cause an increase in electron density and a decrease in electronegativity. In the discharge driven by a triple frequency sawtooth up voltage waveform, due to the rapid decrease of the driving voltage at 0.1 T, the powered sheath expands rapidly, leading to a significant enhancement of electron heating near the powered electrode at 0.1 $T$. The sheath expansion heating of electrons alone is now strong enough to accelerate electrons to high enough energies to cause ionization, so that a maximum of the ionization is observed closed to the expanding sheath edge at the powered electrode at this time. As there are still substantial electric fields in the bulk and close to the collapsing sheath edge at the grounded electrode, ionization appears near the powered sheath expansion, across the bulk region, and near the grounded sheath collapse. Similarly to the dual frequency case, no significant ionization occurs at 0.75 $T$. Although the ionization peak is reduced compared to that in the dual frequency case, it still causes an increase of the electron density, because it occurs over a larger spatial region so that the total, space and time averaged ionization is enhanced. Based on the above discussion, as $N$ increases, there is a significant enhancement in positive electron power absorption at 0.1 $T$ due to the rapid instantaneous drop of the driving voltage waveform. This enhances the ionization. This is the main reason why the electron density increases and the electronegativity decreases.

The changes of the electron power absorption dynamics induced by tailoring the driving voltage waveform affect the high energy tail of the time averaged EEDF as shown in figure 4. In the single frequency case, the spatial distribution of the time averaged EEDF is symmetrical between the powered and grounded electrode. The maximum electron energy is at about 30 eV and the high energy tail of the EEDF is weakly populated. As the number of harmonics $N$ increases to 2 and 3, there are more high energy electrons near the grounded electrode, while the electron energy near the driven side is significantly lower. In the cases of $N$ = 2 and 3, the maximum electron energy in the vicinity of the grounded side even reaches about 38 eV and 40 eV. These results show that using sawtooth up tailored voltage waveforms allows one to generate energetic electrons predominantly and more efficiently at one compared to the other electrode in an otherwise symmetric discharge. This is caused by the effects of the tailored waveform on the electron power absorption dynamics described before, i.e., electrons are accelerated more strongly towards the top electrode than towards the bottom electrode. Consequently, reaction rates such as ionization rates become asymmetric. Therefore, the electron density tends to be higher near the grounded side, as shown in figure 3. These effects can also affect the spatial distribution of the generation of process relevant reactive radicals by electron impact collisions with neutrals.

A DC self-bias voltage is generated as a consequence of the (slope) electrically asymmetric voltage waveforms. In the case of 200 mTorr and Ar/$O_2$ gas ratio of 10/90, the DC self-bias voltage increases form −41.57 V to −36.57 V as the harmonic frequencies $N$ increases from 2 to 3. This phenomenon can still be found at lower $O_2$ ratios. In previous works performed in electropositive gases, as the number of harmonics increases and the difference between the rising and falling slopes of the driving voltage waveform increases further, the DC self-bias voltage was found to be positive for sawtooth up waveforms and to increase as a function of $N$, because the discharge was operated in the α-mode and the ionization maximum occurred during sheath expansion at the powered electrode [31,32]. However, under the discharge conditions studied in this work, the opposite is found, i.e., it is negative and its absolute value decreases as a function of the number of harmonic frequencies $N$. This is caused by the different mode of operation of the plasma in the DA- or hybrid α-DA mode, where the ionization maximum is shifted towards the grounded electrode and occurs during the local sheath collapse. These results of the DC self-bias behavior are similar to those reported in references [34, 59, 60].

### 3.2 Effects of the neutral gas pressure on the discharge characteristics

The spatial and temporal distributions of the electron density, electric field, electron power absorption rate, and total ionization rate are shown in figure 5, respectively, at pressures of 200 mTorr, 300 mTorr, and 500 mTorr, an Ar/$O_2$ mixture of 10/90, and a harmonic number $N$ of 2. Figure 5 shows that increasing the pressure from 200 mTorr to 300 mTorr and 500 mTorr leads to an increase of the peaks of the electron as well as an increase of the electron density in the bulk region. And with the increase of the pressure, the peak of $O^-$ density increases slightly from $6.44\times10^9$ $cm^{-3}$ to $7.06\times10^9$ $cm^{-3}$ and $7.60\times10^9$ $cm^{-3}$. The above reasons lead to a slight decrease of the electronegativity, from 10.96 to 10.71 and 9.91. Moreover, the $O_2(a^1\Delta_g)$ density increases as a function of the gas pressure, which agrees with the observations made in [65]. This is caused by the enhanced low energy (less than 10 eV) electron population at higher pressures, which contributes to more $O_2(a^1\Delta_g)$ generation through the reaction $e+O_2 \rightarrow O_2(a^1\Delta_g)+e$, the threshold of which is 0.0977 eV and the cross section of which has a maximum value at around 8 eV. Furthermore, the $O^-$ density profile gradually moves to the powered side, but the peak of the $O^-$ density is always located closer to the grounded side.

At a pressure of 200 mTorr and as shown in figure 5, the peak of the negative electric field occurs at 0.1 *T* during the sheath collapse phase at the grounded electrode and in the bulk region. There is also a peak of the positive electric field at 0.75 *T* in the bulk. When the pressure is increased to 300 mTorr, the peaks of the negative electric field are located at 0.1 *T* near the expansion phase at the powered electrode and at the grounded electrode during the sheath collapse phase. Moreover, a peak of positive electric field can be observed at 0.75 *T* near the powered electrode during the local sheath collapse phase. However, at 500 mTorr, there are two peaks of the electric field at two different times within one fundamental RF period, at 0.1 *T* and at 0.7 *T*.

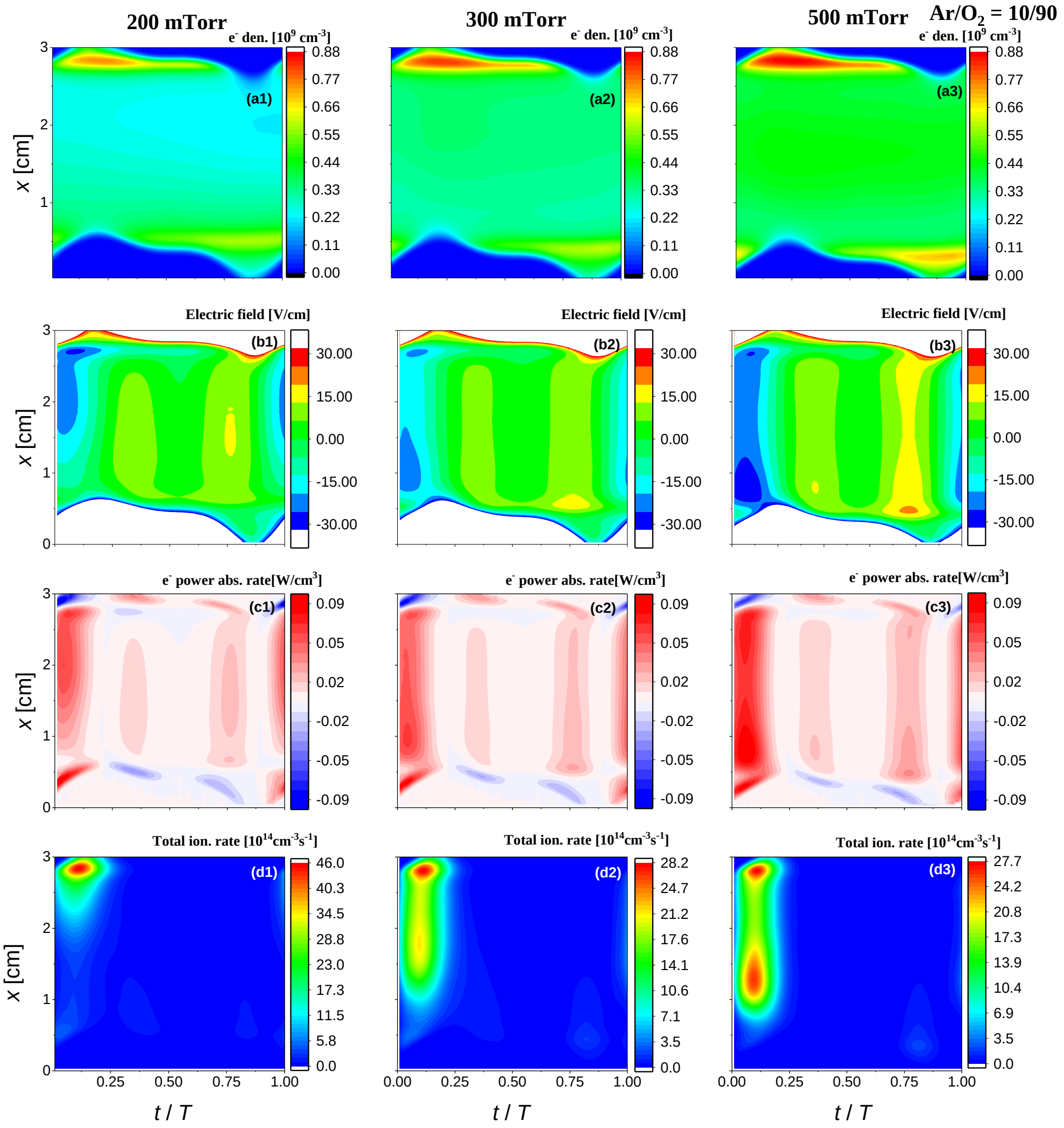

**Figure 5** Spatio-temporal evolutions of the electron density (a1)-(a3), the electric field (b1)-(b3), electron power absorption rate (c1)-(c2) and total ionization rate (d1)-(d3) as a function of pressure (200 mTorr, 300 mTorr, 500 mTorr) for a gas mixture of $Ar/O_2$ = 10/90 sustained by a dual frequency (*N* = 2) sawtooth up voltage waveform. Discharge conditions: gas mixture of $Ar/O_2$ (10/90) with an electrode gap of 3 cm, sustained by a dual frequency voltage waveform with a fundamental frequency of 13.56 MHz and an amplitude of 150 V.

Figures 5 (c) and (d) show that, due to the heating of the electrons by the electric field during sheath expansion at the powered electrode at the expanding sheath edge, due to the drift field in the bulk region, and the ambipolar field near the grounded electrode during the local sheath collapse at 0.1 $T$, the electrons gain enough energy to cause strong ionization at 0.1 $T$ at the different pressures. As the pressure increases, the electron heating is enhanced and the location of the peak electron power absorption spreads towards the expanding sheath edge at the powered electrode. Consequently, the position of maximum ionization also spreads from the grounded to the powered electrode as the pressure increases from 200 mTorr to 500 mTorr. These changes of the position of strong ionization are related to changes of the electron heating dynamics and the decrease of the electron mean free path at higher pressure. The peak ionization rate in the bulk region is attenuated at higher pressure (300 mTorr and 500 mTorr). To explain this phenomenon, the EEDFs at different pressures are shown in figure 6. The proportion of high energy electrons declines due to the increase of the collision frequency with the increasing pressure. This is the main reason for the decrease in the peak of the ionization rate. At 200 mTorr, high energy electrons are predominantly located close to the grounded electrode, and the maximum value of the electron energy is about 40 eV. Meanwhile, low energy electrons tend to be distributed near the driven side. And when the pressure increases to 300 mTorr and 500 mTorr, the maximum value of the electron energy is reduced to about 30 eV, accompanied by the distribution of high energy electrons appearing not only near the grounded side but also in the bulk region. This corroborates to some extent that the ionization rate is stronger in the bulk region at high pressures (300 mTorr and 500 mTorr). As the pressure increases, although the peak of the ionization rate diminishes, the spatial region, where strong ionization occurs, increases significantly, which still leads to an increase in the electron density, as shown in figure 6 (a).

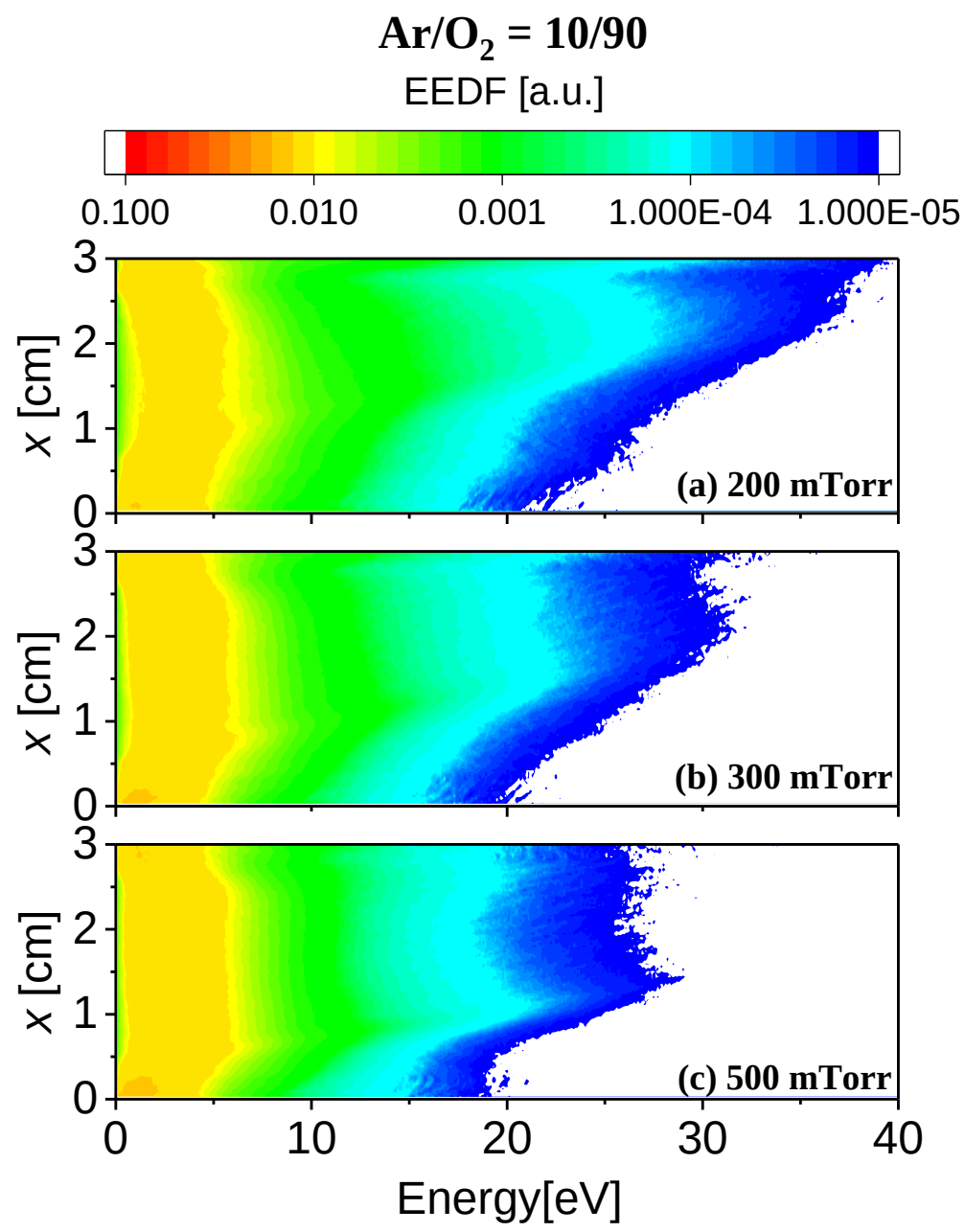

**Figure 6** Fundamental frequency period-averaged and spatially resolved EEDF at (a) 200 mTorr, (b) 300 mTorr, and (c) 500 mTorr. The discharge conditions are the same as those in figure 5.

The magnitude of the self-bias voltage decreases form −41.57 V to −30.25 V and −19.70 V for the $Ar/O_2$ gas ratio of 10/90 as the pressure increases from 200 mTorr to 300 mTorr, 500 mTorr. Figure 5 (d) shows that at lower pressure (200 mTorr), the peaks of the ionization appear only near the grounded electrode. When the pressure increases to 300 mTorr and 500 mTorr, the peaks of the ionization not only appear near the grounded electrode, but also in the bulk region. This leads to an enhanced symmetry of the plasma and, thus, reduces the absolute value of the DC self-bias.

### 3.3 Effects of the gas mixture on the discharge characteristics

Figure 7 show the temporal and spatial evolution of the electron, the electric field, the rate of electron power absorption, and the total ionization rate at a pressure of 200 mTorr, a harmonic number $N$ of 2, and for Ar/$O_2$ gas mixture ratios of 90/10, 50/50, and 10/90. As the percentage of $O_2$ increases from 10% to 50% and 90%, there is an increase in electronegativity from 3.6 to 8.3 and 10.97. In figure 7, the electron density in the bulk region declines as the percentage of $O_2$ increases, while peaks of the electron density at the sheath edge become more apparent and the negative ion density increases significantly. Moreover, the peak of $O^-$ density increases from $4.58\times10^9$ cm$^{-3}$ to $5.94\times10^9$ cm$^{-3}$ and $6.44\times10^9$ cm$^{-3}$. And the increase of the $O_2$ content leads to a higher $O_2(a^1\Delta_g)$ density. Besides, with the increase of $O_2$ percentage, the electron density tends to be highest near the grounded electrode, and the same trend can also be found for the ions.

Figure 7 (b) shows that the electronegativity increases with increasing $O_2$ percentage, while the electron density in the bulk region decreases, the conductivity decreases, and the drift and ambipolar fields in the bulk region are significantly enhanced. Taking the electric field at 0.1 $T$ and at the position of 0.2 cm as an example, the electric field changes from -7.8 V/cm to -10.6 V/cm and -19.3 V/cm when the $O_2$ percentage increases from 10% to 50% and 90%. According to figure 11 (b), electrons are mainly accelerated by sheath expansion heating and the electric field in the bulk region at an $O_2$ percentage of 10%, but sheath expanding heating dominates. When the $O_2$ percentage increases to 50%, the heating in the bulk region is enhanced. When the $O_2$ percentage is further increased to 90%, the heating in the bulk region is comparable to that caused by the sheath expansion, i.e., the DA mode is enhanced by increasing the $O_2$ admixture. Changes of the electron power absorption dynamics directly affect the electron energy distribution. Figure 8 shows a plot of the spatially resolved and time averaged electron energy distribution as a function of the $O_2$ admixture. As the $O_2$ percentage increases from 10% to 90%, the proportion of high energy electrons increases significantly, and the maximum electron energy increases from about 24 eV to 40 eV. High-energy electrons are mostly located in the bulk region and near the grounded electrode at Ar/$O_2$ gas mixture ratios of 90/10 and 50/50. These changes of the EEDF are directly linked to the changes of the electron power absorption dynamics discussed before and are highly process relevant, since they affect the generation of ions and neutral species. Its effects on the ionization dynamics are illustrated by figure 7 (d). The ionization mainly occurs at 0.1 $T$, when the absolute value of the electron conduction current is maximum. At a relatively low $O_2$ percentage (10%), the ionization is strong in the vicinity of the expanding sheath at the bottom powered electrode, in the bulk region and close to the collapsing sheath at the grounded electrode. The discharge is operated in a hybrid α- and DA-mode. When the $O_2$ percentage increases, the ionization at the bottom electrode during the sheath expansion phase and in the bulk region becomes gradually weaker compared to that at the collapsing sheath edge at the grounded electrode, which indicates a heating mode transition into the DA-mode.

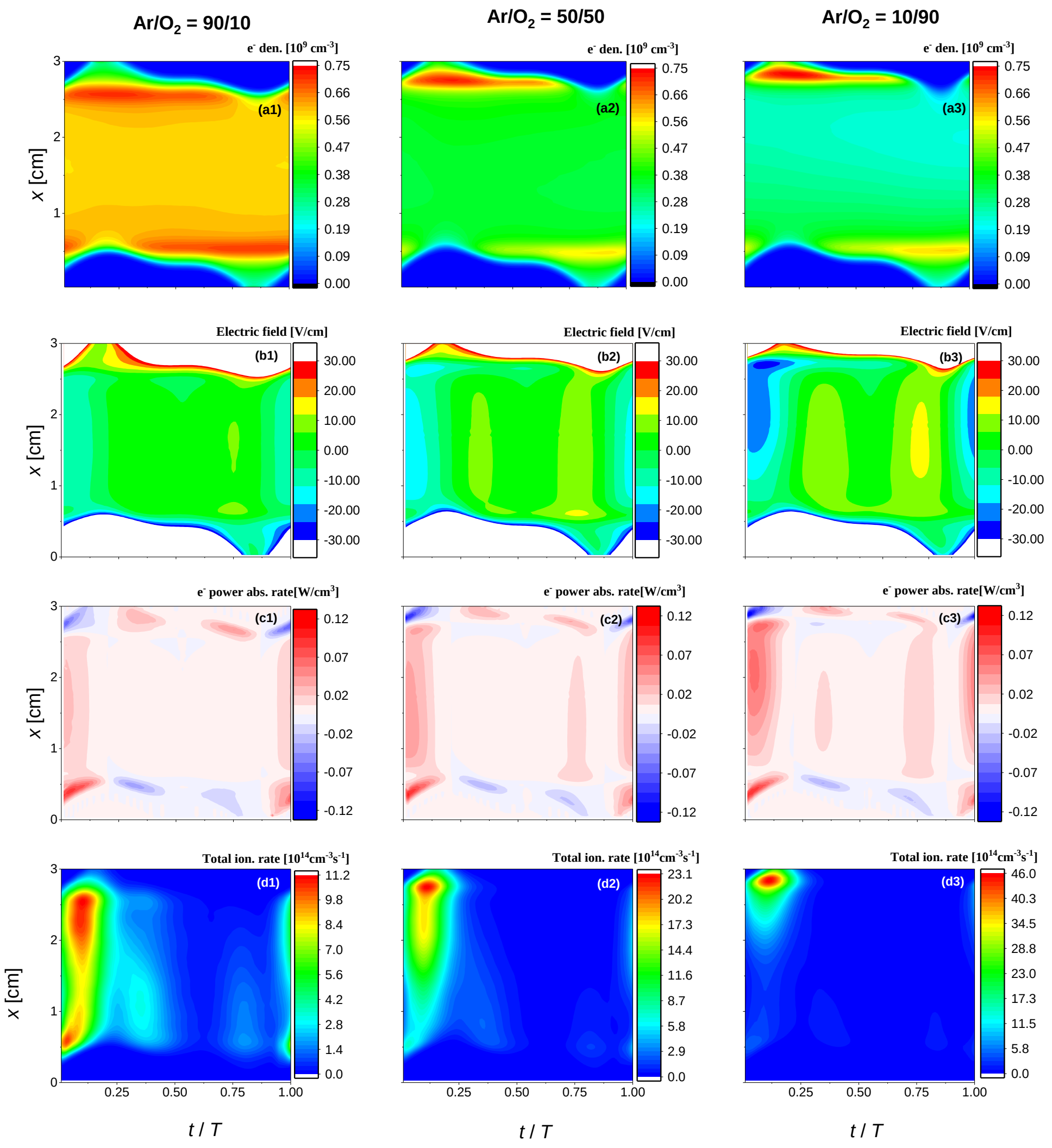

**Figure 7** Spatio-temporal evolutions of the electron density (a1)-(a3), the electric field (b1)-(b3), electron power absorption rate (c1)-(c3) and total ionization rate (d1)-(d3) at 200 mTorr for different gas mixture ratios of Ar/ $O_2$ (90/10, 50/50, 10/90) in a dual frequency ($N$ = 2) discharge. Discharge conditions: The electrode gap is 3 cm and the discharge sustained by dual frequency excitation with a fundamental frequency of 13.56 MHz and 150 V amplitude.

As a function of the $O_2$ percentage, the magnitude of the DC self-bias voltage increases. This is caused by the following mechanism: According to figure 7 (d), at a low $O_2$ admixture, the discharge is operated in a hybrid α-/DA-mode, i.e., ionization appears near the grounded electrode, powered electrode, and in the bulk region. Ions are, thus, generated in the bulk region and near the sheath boundary at both electrodes, and the ion densities at both electrodes do not differ much. With the increase of the $O_2$ admixture, the DA mode dominates, and the ionization mainly happens in the vicinity of the grounded electrode. Consequently, more ions are generated near the grounded electrode, which increases the local ion density relative to that at the powered electrode. This makes the plasma asymmetric and causes a stronger DC self-bias.

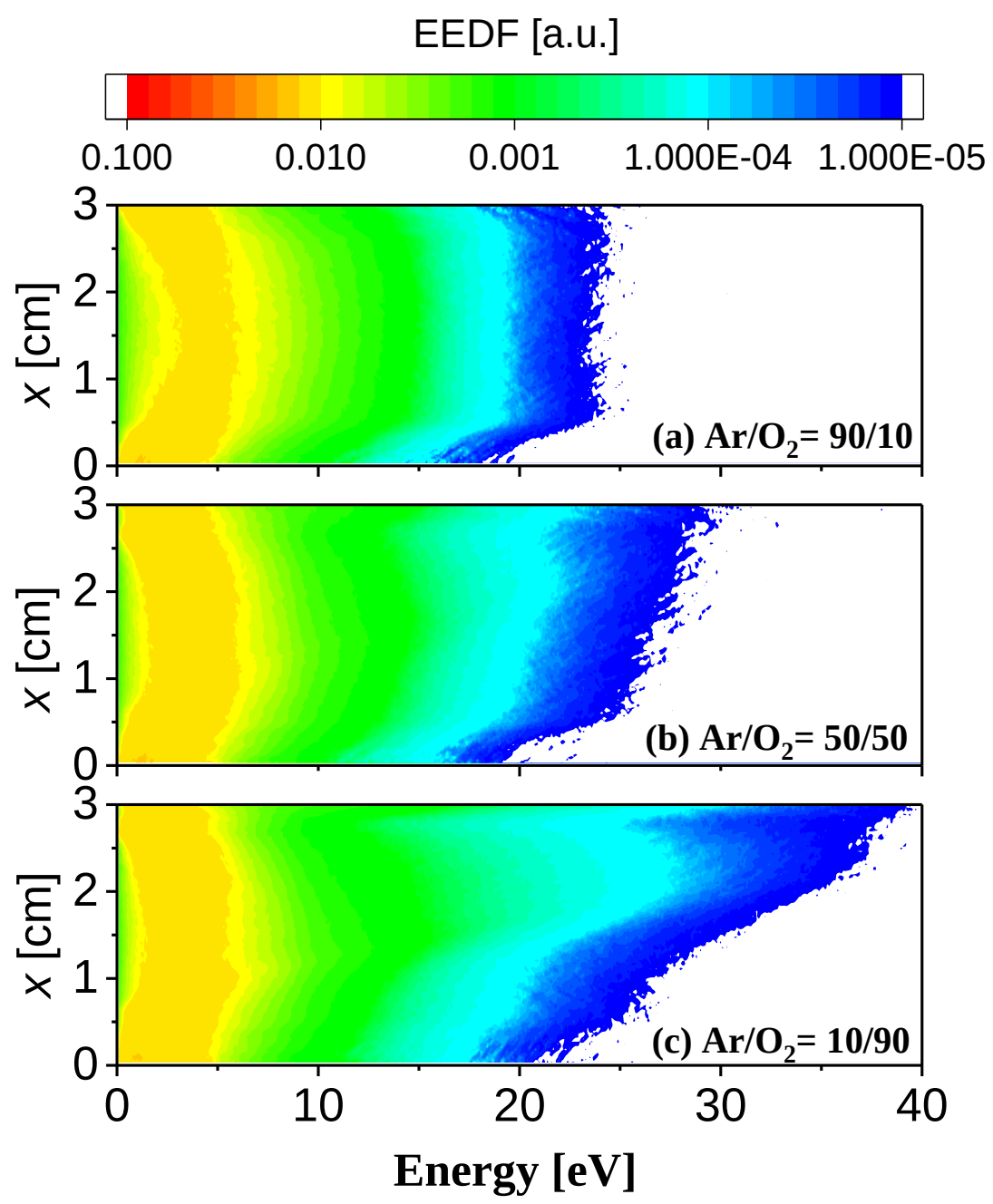

**Figure 8** Space resolved and fundamental frequency period-averaged EEDF at Ar/$O_2$ gas mixture ratios of (a) 90/10, (b) 50/50, and (c) 10/90. The discharge conditions are the same as those in figure 7

### 3.4 The effects of external control parameters on the transport of charged and neutral particles

Figure 9 illustrates the space and time-averaged charged particle densities and electronegativity $\eta$ as a function of the gas mixture ratio, pressure, and of the number of consecutive harmonic frequencies $N$. At the same $O_2$ percentage, there is an enhancement of the electron, negative, and positive ion densities as a function of $N$ and, meanwhile, a decrease of the electronegativity can be observed. Although the spatio-temporal region of high ionization rate is reduced (see figure 3 (d)), the larger peak of the ionization rate still results in an increase in positive ion densities when the harmonic frequency number increases from 1 to 2. In the case of 3 consecutive harmonics, the peak of the ionization rate declines, but the region of high ionization is enlarged compared to $N = 2$, which causes a further increase of the positive ion density. Since the trends of the densities of electrons, $O_2$ associated positive ions, $Ar^+$, $O^-$, and electronegativity is the same with the decreasing Ar percentage for all values of $N$, we choose $N = 2$ as an example to discuss the effect of the $O_2$ percentage on the charged particle densities and the electronegativity. With the increase of the electronegative $O_2$ admixture, the densities of $O_2$-associated positive and negative ions increase, the electron density gradually decreases, the density of $Ar^+$ gradually decreases, and the electronegativity increases. As shown in figure 7 (d), the peaks of the ionization rate gradually get stronger as the Ar percentage decreases, which leads to a significant increase of the $O_2$-associated positive ion density. And due to the increase of the $O_2$ admixture, the $O_2$ attachment reaction is also enhanced, which results in a decrease of the electron density and an increase of the $O^-$ density. Therefore, the electronegativity increases and there is a change of the discharge mode from a weakly electronegative discharge operated in a hybrid α-/DA-mode to a strongly electronegative discharge, where the DA-mode dominates.

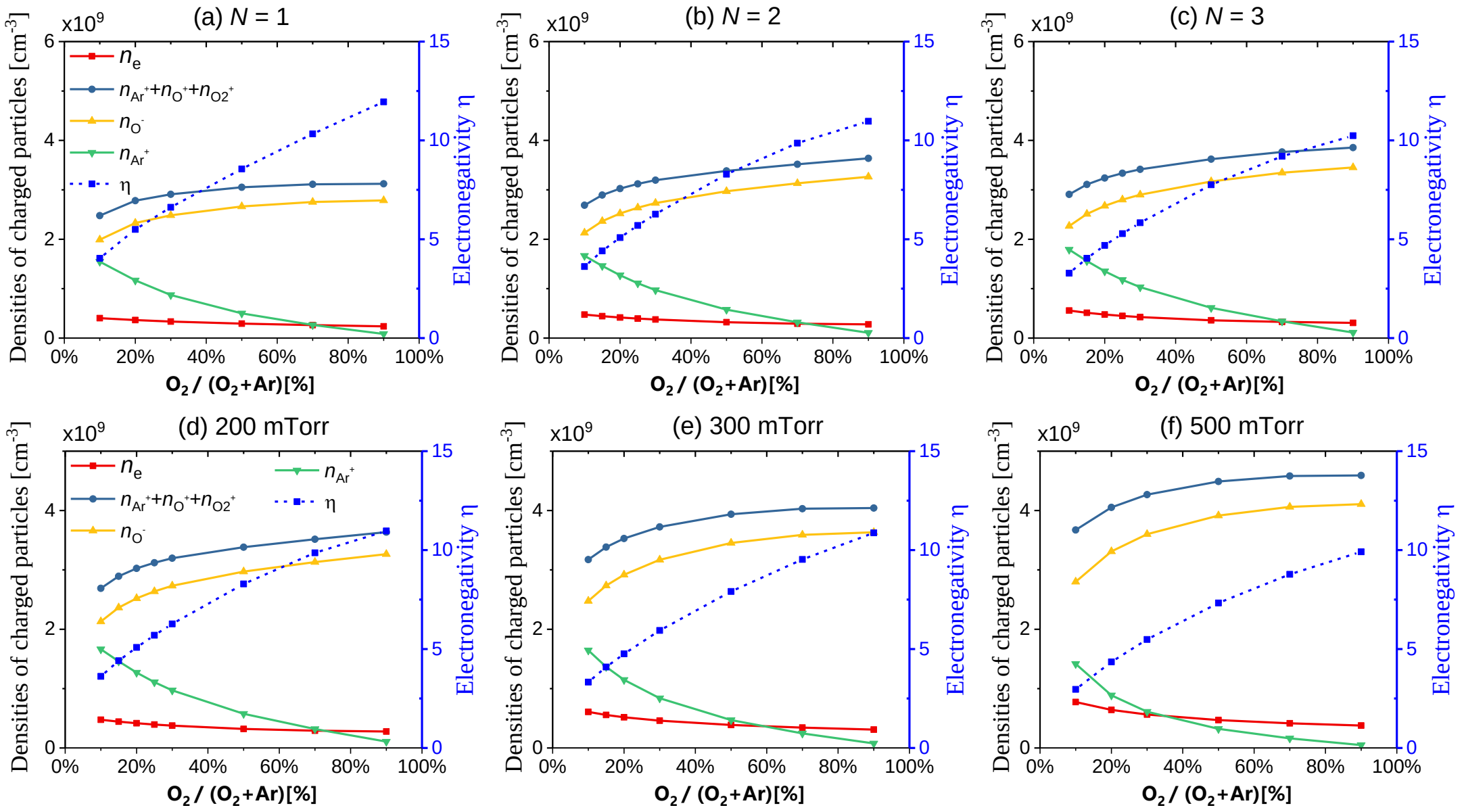

**Figure 9** Space and fundamental frequency period averaged electron density (red line), total positive ion density (black line, the sum of $Ar^+$ density, $O^+$ density, and $O_2^+$ density), $O^-$ density (orange line), $Ar^+$ density (green line), and electronegativity (blue line) η. First row: (a) $N$ = 1, (b) $N$ = 2, (c) $N$ = 3 in the case of 200 mTorr, second row: (d) 200 mTorr, (e) 300 mTorr, (f) 500 mTorr at $N$ = 2. Discharge conditions: $Ar/O_2$ (90/10~10/90) with an electrode gap of 3 cm, sustained by single or multiple frequency excitation with a fundamental frequency of 13.56 MHz and an amplitude of 150.

As the pressure increases, the electron, negative ion, and positive ion densities increase, and there is a slight reduction of the electronegativity. According to figure 5 (d), when the pressure is gradually increased from 200 mTorr to 500 mTorr, the ionization peak is attenuated, but the spatio-temporal region of high ionization is significantly enlarged, which causes the total ionization to increase, so that the positive ion and electron densities increase. Due to the increase of the background gas density, more electron attachment happens, resulting in a rise of the negative ion density. Although the rate of the $O^-$ loss reaction (such as the reaction of $O^-$ with $O_2(a^1\Delta_g)$) increases, it plays a minor role. The increase of the electron density in the bulk region leads to a decrease of the electronegativity. The trends of the individual particle densities as a function of the gas mixture at 300 mTorr and 500 mTorr are similar to those shown in figure 9 for 200 mTorr.

Figure 10 shows the period averaged neutral fluxes of $Ar^*$, $O(^3P)$, and $O(^1D)$ at the driven electrode (first column) and at the grounded electrode (second column) for different gas mixture ratios and numbers of harmonic driving frequencies $N$ = 1, 2 and 3, respectively. At the powered electrode, the fluxes of $O(^3P)$, $O(^1D)$ and $Ar^*$ decrease as a function of $N$ for high $O_2$ admixtures, while a different behavior is found at the grounded electrode. The $Ar^*$ flux at the grounded electrode increases as a function of $N$ for high $O_2$ percentage. Meanwhile, the $O(^3P)$ flux declines, the flux of $O(^1D)$ increases and then decreases slightly. At low $O_2$ percentage (< 30%) and at the driven electrode, the $Ar^*$ flux decreases as a function of $N$, while the $O(^3P)$ flux shows an increasing trend. The flux of $O(^1D)$ reaches a minimum value in the case of dual frequency excitation and a maximum value at $N$ = 3. On the grounded electrode and for low $O_2$ admixtures, the fluxes of the three neutral particles gradually increase as a function of $N$.

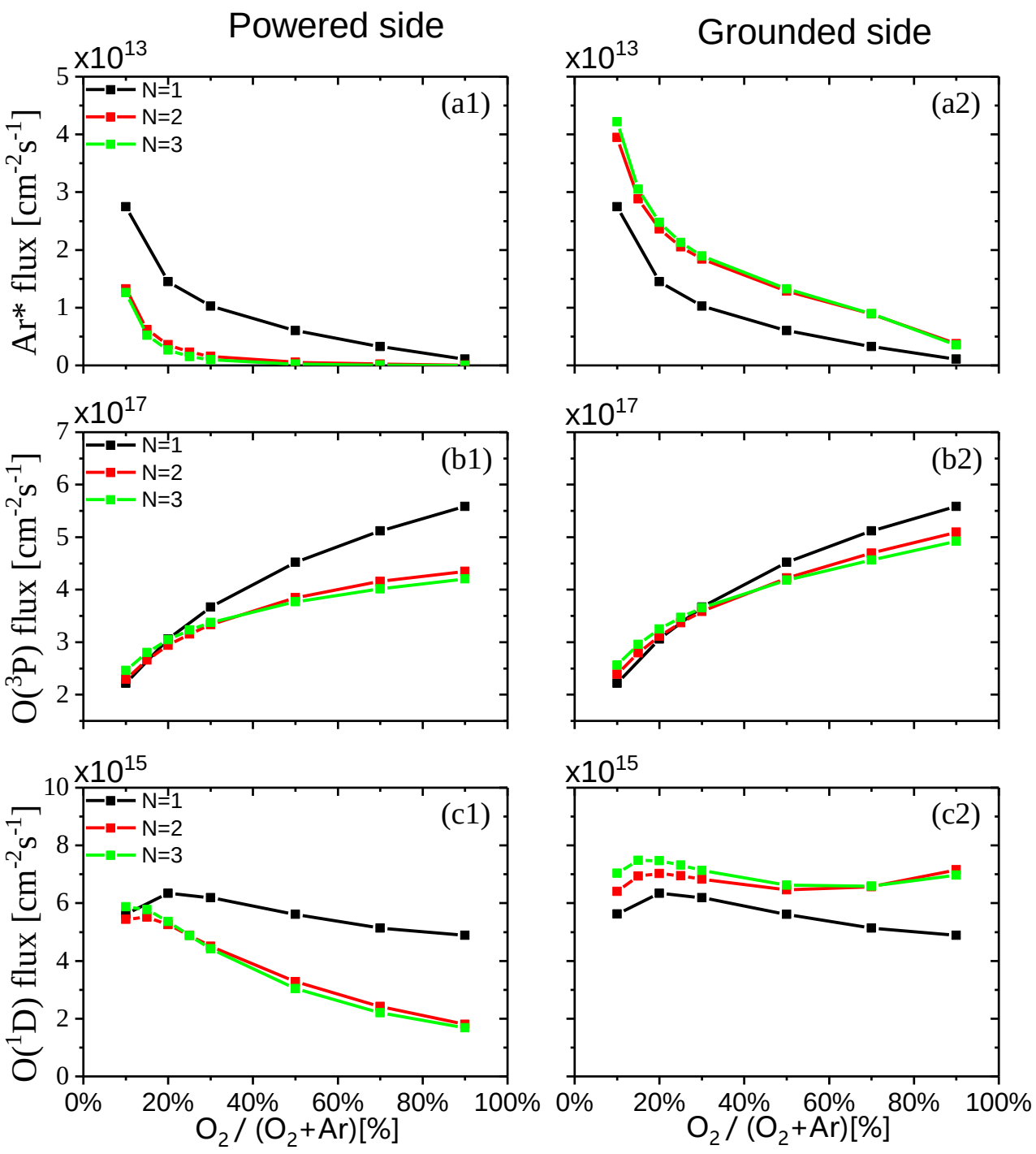

**Figure 10** Fundamental frequency period averaged fluxes of (a1)-(a2) $Ar^*$, (b1-(b2) $O(^3P)$ and (c1)-(c2) $O(^1D)$, for $N$ = 1 (black line), $N$ = 2 (red line), $N$ = 3 (green line) at the powered electrode and at the grounded electrode. Discharge conditions: $Ar/O_2$ (90/10~10/90) at 200 mTorr with an electrode gap of 3 cm sustained by single or multiple frequency excitation with a fundamental frequency of 13.56 MHz and 150 V voltage amplitude.

Firstly, for an exemplary $O_2$ percentage of 10%, the variations of the three neutral fluxes at different numbers of harmonic frequencies $N$ will be discussed. Figure 11 show the spatio-temporal distribution of the generation and loss rate for $Ar^*$, $O(^3P)$, and $O(^1D)$ at different numbers of harmonic frequencies $N$. Based on figures 11 (a) and (b), the generation rate of $Ar^*$ shifts to the grounded side as the number of harmonic frequencies increases. This is related to the asymmetric discharge characteristics caused by using asymmetric driving voltage waveforms. Ultimately this is caused by the enhancement of the high energy tail of the EEDF as close to the grounded electrode and its depletion at the powered electrode, induced by increasing $N$ for a sawtooth up voltage waveform, as a consequence of the effects of this tailored voltage waveform shape on the electron power absorption dynamics. For this reason, more electrons are present at the grounded electrode with energies above the threshold for $Ar^*$ excitation of about 11.56 eV. Specifically, on the grounded electrode, the peak value of the $Ar^*$ generation term significantly increases from $22.7\times10^{14}$ $cm^{-3}s^{-1}$ to $31.0\times10^{14}$ $cm^{-3}s^{-1}$ and $39.6\times10^{14}$ $cm^{-3}s^{-1}$, and the peak of the generation rate is located closer to the grounded electrode. The reaction $O_2 + Ar^* \rightarrow O(^3P) + O(^3P) + Ar$ mentioned in our previous work as R47 [23] is the main loss channel of $Ar^*$. Since this reaction rate coefficient is a fixed value ($2.1\times10^{-10}$), it is independent of the electron dynamics, and only the distribution of the $Ar^*$ density may affect the spatio-temporal distribution of this reaction rate. From figure 11 (b), as $N$ increases, the changes in both the peak and the distribution of the loss rate are not significant. Therefore, these factors result in an increase of the $Ar^*$ flux at the grounded electrode. And a decrease of the $Ar^*$ flux on the driven side as the number of harmonic frequencies increases can also be observed, where the number of energetic electrons decreases as a function of $N$.

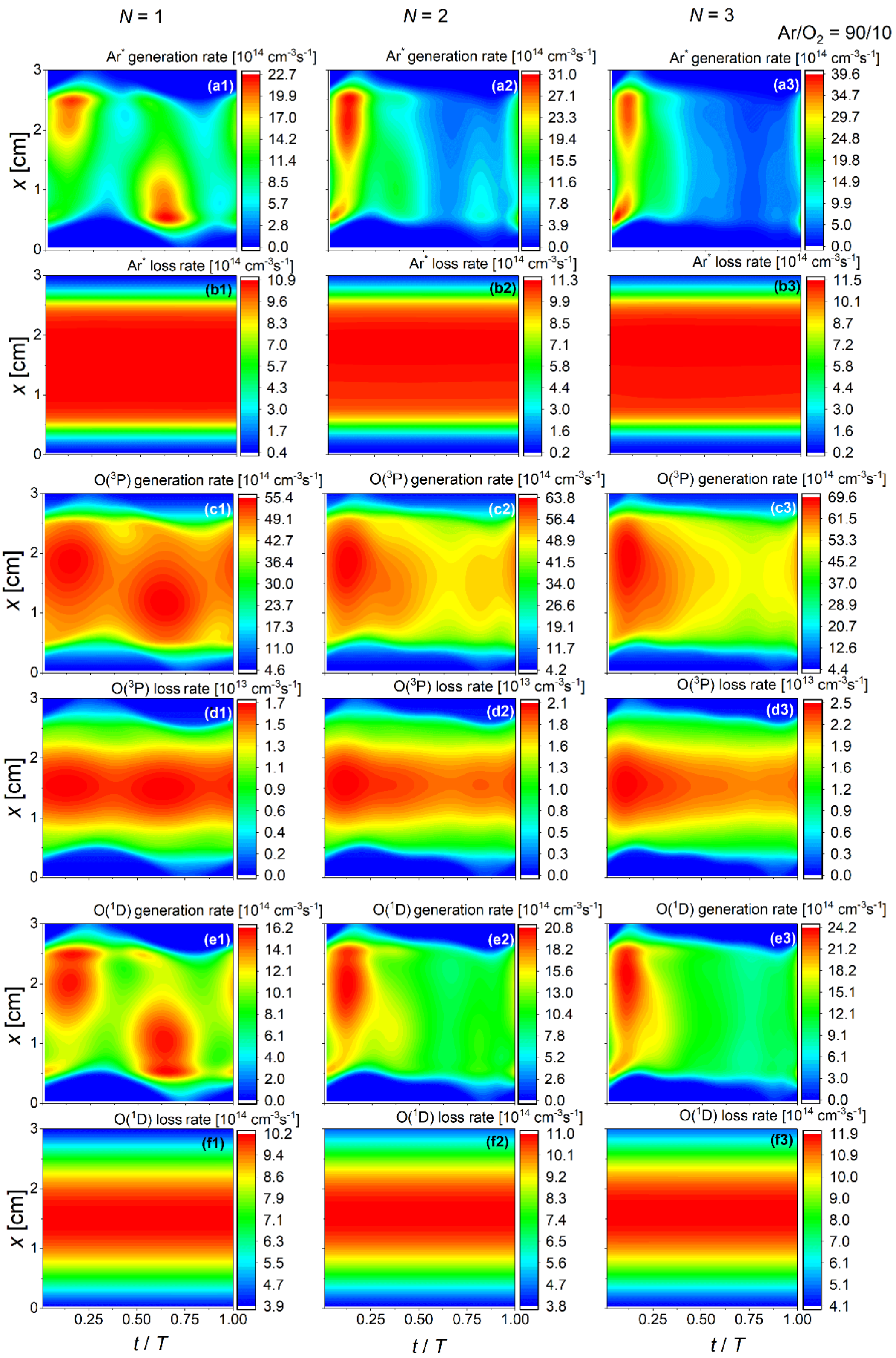

**Figure 11** Spatio-temporal evolutions of (a1)-(a3) the $Ar^*$ generation rate, (b1)-(b3) the $Ar^*$ loss rate, (c1)-(c3) the $O(^3P)$ generation rate, (d1)-(d3) the $O(^3P)$ loss rate, (e1)-(e3) the $O(^1D)$ generation rate, and (f1)-(f3) the $O(^1D)$ loss rate for a gas mixture ratio of $Ar/O_2$ = 90/10 at (a) $N$ = 1, (b) $N$ = 2, (c) $N$ = 3. The discharge conditions are the same as those in figure 15

The $O(^3P)$ flux shows an increasing trend on both the grounded and driven sides as the number of harmonic frequencies increases at a low admixture of $O_2$ to Ar. According to the reaction set in this work, the generation of $O(^3P)$ is dominated by the reactions $e + O_2 \rightarrow O(^3P) + O(^1D) + e$ and $O_2 + Ar^* \rightarrow O(^3P) + O(^3P) + Ar$ mentioned in our previous work as R18 and R47 [23]. The dissociative excitation threshold for R18 is only 8.4 eV, the reaction rate coefficient of R47 is a fixed value and its rate is related to the densities of $O_2$ and $Ar^*$. This is because the reaction R47 is independent of the electron density distribution and electron energy distribution, and the distribution of the neutrals is almost unaffected by the discharge asymmetry within a fundamental frequency period. The region of high generation rate is still larger at different *N*, compared to the generation rate of $Ar^*$ and $O(^1D)$ (as will be discussed later). In addition, the magnitude of the loss rate is much smaller than the generation rate. This means that the diffusion term may play an important role compared to the loss rate. Nevertheless, the primary loss reactions for $O(^3P)$ can be identified. When the $O_2$ gas admixture is low, R22 ($e + O(^3P) \rightarrow O(^1D) + e$), and R48 ($O(^3P) + Ar^* \rightarrow O(^1D) + Ar$) dominate the loss processes. However, as the $O_2$ percentage increases, the reaction R41 ($O^- + O(^3P) \rightarrow O_2 + e$) becomes more significant due to the substantial decrease of the Ar-related particle density. From figure 11 (c), with the increasing number of harmonic frequencies, the peak value of the $O(^3P)$ generation term significantly increases from $55.4\times10^{14}$ cm$^{-3}$s$^{-1}$ to $63.8\times10^{14}$ cm$^{-3}$s$^{-1}$ and $69.6\times10^{14}$ cm$^{-3}$s$^{-1}$, while the spatial and temporal region of high generation rates remains approximately the same. This is different from the generation rate of $Ar^*$, which is predominantly located close to the top grounded electrode for high values of *N*. This difference is ultimately caused by the lower electron threshold energy for the generation of $O(^3P)$ compared to $Ar^*$ and the fact that this lower energy part of the EEDF is less strongly affected by VWT as compared to its higher energy part. More generation of $O(^3P)$ means that more $O(^3P)$ can diffuse towards the electrode. Therefore, an increasing trend of the $O(^3P)$ flux for low $O_2$ admixtures at both powered and grounded electrode surface can be observed.

The generation of $O(^1D)$ is mainly related to the reaction $e + O_2 \rightarrow O(^3P) + O(^1D) + e$ (R18), which can be affected by the electron dynamics. According to figure 11 (e), compared to single frequency operation, the generation rate at the cases of $N = 2$ and 3 is only one peak at around 0.1 *T*, and the peak of the generation rate is significantly larger near the grounded electrode. The main loss reactions for $O(^1D)$ are R49 ($O(^1D) + Ar \rightarrow O(^3P) + Ar$), and R53 ($O(^1D) + O_2 \rightarrow O(^3P) + O_2$) mentioned in our previous work [23]. It can be seen that the loss reaction is not directly related to the electron dynamics, and therefore the loss rate is not affected by the electrically asymmetric discharge properties. At different *N*, the peak and spatio-temporal evolution of the loss rate remains almost unchanged. Therefore, more generation with almost constant loss, which causes more diffusion of the $O(^1D)$ to the electrode, results in an increase of the $O(^1D)$ flux near the grounded electrode. On the driven electrode, because of the asymmetric discharge, compared to the result obtained for single frequency operation, the region of strong generation rate near the driven electrode shrinks at $N = 2$, leading to a decrease in flux at the driven electrode. At $N = 3$, the peak value of the generation rate on the driven electrode is enhanced, and the change in the loss rate is not significant, further resulting in an increase in $O(^1D)$ flux compared to the result at $N = 2$.

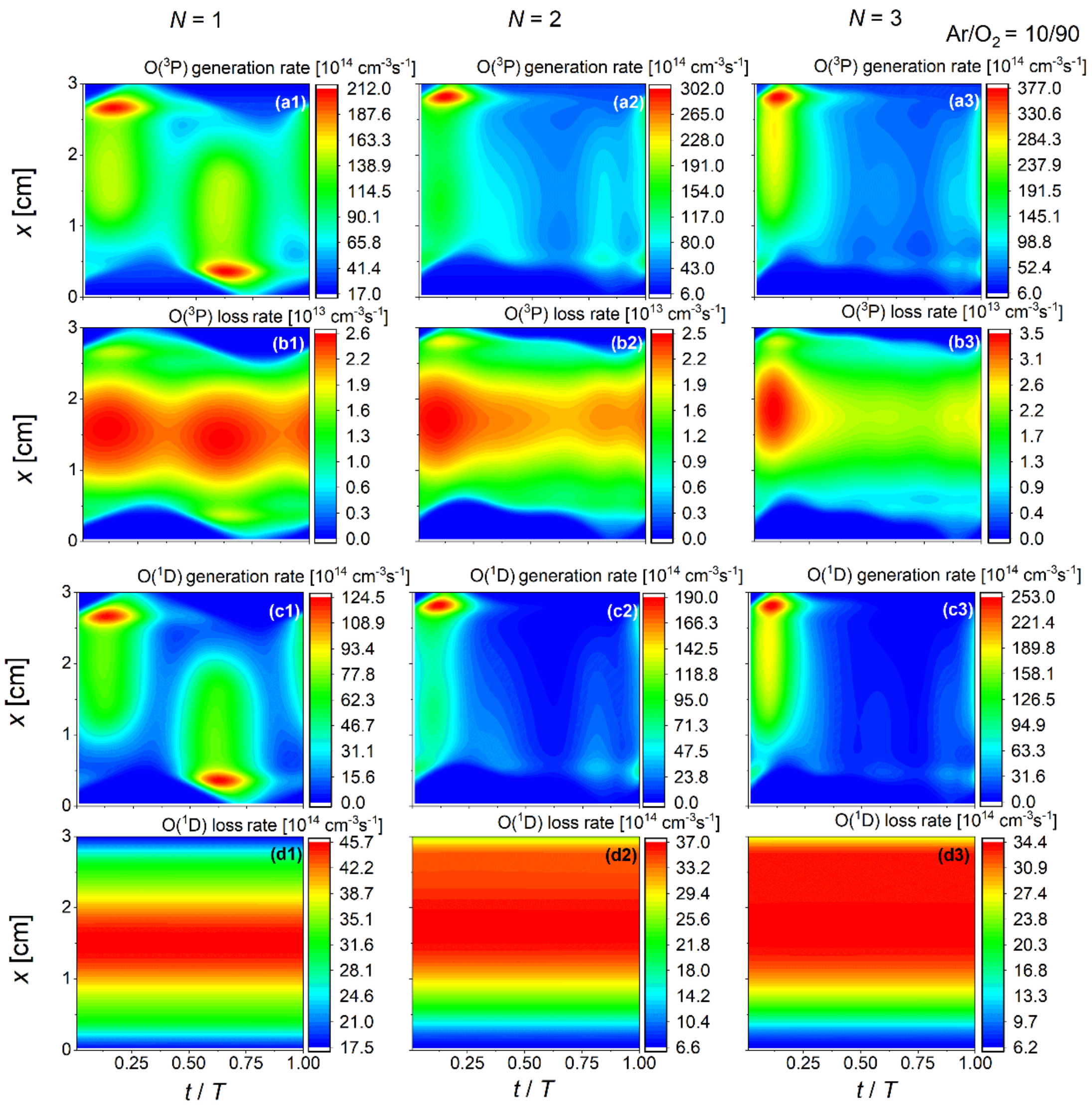

**Figure 12** Spatio-temporal evolutions of (a1)-(a3) the $O(^3P)$ generation rate, (b1)-(b3) the $O(^3P)$ loss rate, (c1)-(c3) the $O(^1D)$ generation rate, and (d1)-(d3) the $O(^1D)$ loss rate for a gas mixture ratio of $Ar/O_2$ = 10/90 at (a) $N$ = 1, (b) $N$ = 2, (c) $N$ = 3. The discharge conditions are the same as those in figure 15.

To analyze the variations in the flux of $Ar^*$, $O(^3P)$, and $O(^1D)$ at different number of harmonic frequencies under higher $O_2$ percentage, this section will take an $O_2$ percentage of 90% as an example. Since the flux of $Ar^*$ varies consistently as a function of the harmonic frequencies number at different oxygen ratios, and the physical mechanisms are the same, it will not be elaborated further here. Figure 12 show the spatio-temporal evolution of the generation and loss rates of $O(^3P)$ and $O(^1D)$. From figures 12 (a) and (b), as the number of harmonic frequencies increases, the peak of the generation term of $O(^3P)$ changes from being pronounced at both electrodes to occurring only near the grounded electrode. Besides the increase in the peak of the generation rate, the size of the spatial and temporal region of high generation decreases. The loss rate still plays a minor role due to its smaller magnitude than that of the generation rate. The effect of reduction of the size of the spatial and temporal region of high generation is more pronounced, leading to a decrease in the flux of $O(^3P)$ at the powered electrode. Moreover, since the generation source term is higher near the grounded electrode, the flux of $O(^3P)$ at the grounded electrode is higher than that at the driven electrode in the case of $N = 2$ and 3. Besides, due to the reduction in the spatial and temporal region of strong generation, the flux of $O(^3P)$ at the grounded electrode also decreases as the number of harmonic frequencies increases.

According to figures 12 (b), and (d), there are some differences between the spatial and temporal distributions of the loss rates of $O(^1D)$ and $O(^3P)$. The larger values of the loss rate of $O(^1D)$ gradually approach the grounded electrode as the number of harmonic frequencies increases, whereas the loss term for $O(^3P)$ is mainly concentrated in the bulk region and plays a minor role due to its smaller magnitude compared to the generation rate. As shown in figure 12, with the increase of the number of harmonic frequencies, the peak of the generation rate of $O(^1D)$ significantly increases, and the location of the peak changes, from being observed at both electrodes to being present only close to the grounded side. Consequently, more $O(^1D)$ is generated near the grounded side and the flux at the grounded electrode for $N = 2$ and $N = 3$ is higher than that for $N = 1$. Additionally, by comparing the loss rates for $N = 1$, 2 and 3, the peak of the loss term for $N = 2$ and 3 is closer to the grounded electrode. The loss of $O(^1D)$ is mainly related to R49 ($O(^1D) + Ar \rightarrow O(^3P) + Ar$), and R53 ($O(^1D) + O_2 \rightarrow O(^3P) + O_2$). The rate coefficients for these two reactions are constant values, and the densities of the background gases Ar and $O_2$ are uniformly distributed, so they are mainly affected by the $O(^1D)$ density distribution. And from the $O(^1D)$ generation rate, the $O(^1D)$ generation rate is higher at the grounded side than that near the driven electrode. Therefore, more $O(^1D)$ is generated near the grounded side, which further moves the peak of the loss rate to a region near the grounded side. Compared to the loss rate at $N = 2$, the peak of the loss rate is closer to the grounded electrode in the case of $N = 3$. This might partly explain why the flux of $O(^1D)$ is a little bit lower at $N = 3$ compared to $N = 2$ at the grounded electrode. Near the powered electrode, due to the asymmetry of the discharge, the size of the spatial and temporal region of high generation rate is significantly reduced at $N = 2$ and $N = 3$, and the peaks are much lower than in the single-frequency scenario, which might cause a reduction of the $O(^1D)$ flux at the powered electrode as the number of harmonic frequencies increases.

In the presence of a slope asymmetric driving voltage waveform, the flux of oxygen-related neutrals to the electrodes can be controlled by VWT. Especially when the $O_2$ admixture and, thus, the electronegativity is low, the flux of oxygen-related neutrals tends to increase with the increase of the number of harmonic frequencies. However, when the admixture of electronegative oxygen increases, this increase is attenuated. This is mainly because the increased electronegativity makes the discharge more localized and the generation rate and loss rate differ greatly from the scenarios with weaker electronegativity. Regarding the flux of Ar-related neutrals, there is a significant difference between the two electrodes. Due to the electrically asymmetry effect, the peak of the generation term moves towards the grounded electrode. Compared to single-frequency discharges, the $Ar^*$ flux to the powered side significantly decreases. Meanwhile, this phenomenon also results in higher $Ar^*$ fluxes to the grounded side compared to the single frequency case.

In addition, at the same number of harmonic frequencies, the flux of $O(^3P)$ increases with the increase in the $O_2$ admixture and the flux of $Ar^*$ decreases. This is mainly due to the increased density $O_2$ and the decreased proportion of Ar. Previously, under single-frequency conditions, it was observed that the flux of $O(^1D)$ initially increased and then decreased with the increase of the $O_2$ admixture, as shown in figure 10(c), with the peak occurring around the $O_2$ admixture of 20%. This phenomenon was addressed in previous studies [21, 23]. The initial increase of the $O(^1D)$ flux is due to an increase of the background gas density of $O_2$, while the subsequent decrease is mainly due to a reduction of the size of the region when/where a high generation rate occurs [23]. In the case of discharges driven by slope asymmetric voltage waveforms, the same trend of the $O(^1D)$ flux can be observed. However, in the case of $N = 2$ and $N = 3$, when the $O_2$ admixture is high (70%-90%), the flux of $O(^1D)$ exhibits an increasing trend at the grounded electrode, which is different from results obtained in the single frequency case. To better analyze the changes in $O(^1D)$ flux with increasing $O_2$ admixture, figure 13 shows the spatio-temporal evolution of the generation and loss rates of $O(^1D)$ when the number of harmonic frequencies is $N = 2$ and the pressure is at 200 mTorr, for $Ar/O_2$ gas mixtures of 90/10, 80/20, 50/50, and 10/90. As the $O_2$ admixture increases from 10% to 20%, 50%, and 90%, the peak of the generation rate of $O(^1D)$ significantly increases from $20.8\times10^{14}$ $cm^{-3}s^{-1}$ to $37.7\times10^{14}$ $cm^{-3}s^{-1}$, $98.5\times10^{14}$ $cm^{-3}s^{-1}$, and $190.0\times10^{14}$ $cm^{-3}s^{-1}$, respectively. Meanwhile, the loss term of $O(^1D)$ also changes, with its peak increasing from $11.0\times10^{14}$ $cm^{-3}s^{-1}$ to $18.1\times10^{14}$ $cm^{-3}s^{-1}$, $29.2\times10^{14}$ $cm^{-3}s^{-1}$, and $37.0\times10^{14}$ $cm^{-3}s^{-1}$. Figure 13 also shows that, for an $O_2$ admixture of 20%, characteristics of a localized discharge and spatial asymmetries start to emerge compared to a lower $O_2$ admixture of 10%. For an $O_2$ admixture of 90%, the generation and loss rates of $O(^1D)$ are significantly influenced by these effects, with peaks mainly concentrated at the grounded electrode. This is the reason for the different $O(^1D)$ fluxes at the two electrodes. When the $O_2$ ratio increases from 10% to 20%, the generation rate increases, especially in the bulk region, which further leads to an increase of the $O(^1D)$ flux. However, due to the reduced time and spatial region when/where a high generation rate occurs (figure 13 (a)), particularly near the powered side, the $O(^1D)$ flux to the driven electrode side decreases. Conversely, on the grounded electrode, due to the asymmetry effect, the peak of the generation rate of $O(^1D)$ is enhanced, resulting in an increase of the $O(^1D)$ flux on this side. As the $O_2$ content continues to increase, the asymmetry and localization become more pronounced, with the peak generation rate of $O(^1D)$ and the diffusion term on the powered electrode side remaining almost unchanged, and the loss rate near the powered side decreases, leading to a continuous decline in $O(^1D)$ flux to the powered electrode. For instance, for an $O_2$ admixture of 90%, the size of the time and spatial region of high $O(^1D)$ generation rate (figure 13 (a4)) significantly decrease, with the peak generation rate localized near the grounded sheath collapse phase. The peak of the generation rate is twice higher than that at 50% $O_2$ admixture, while the loss rate increases by only 50%, causing an increase of the $O(^1D)$ flux to the grounded electrode.

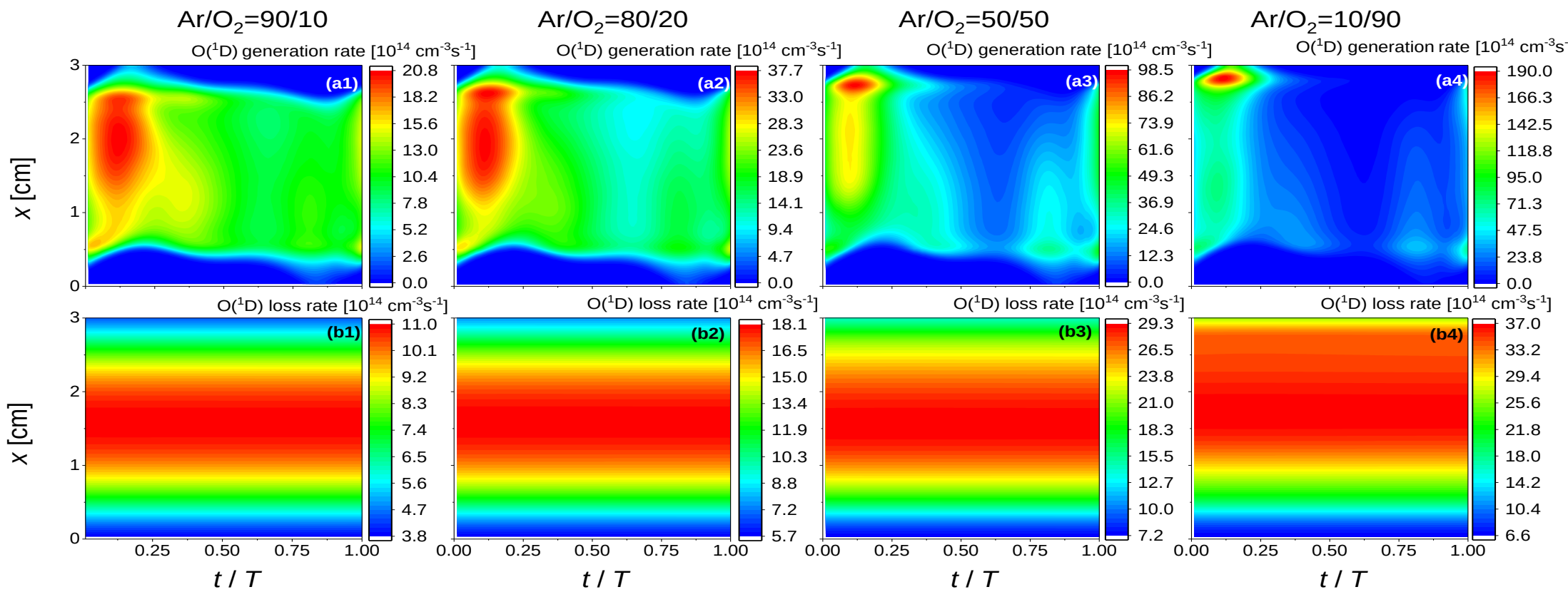

**Figure 13** Spatio-temporal evolutions of (a1)-(a4) the $O(^1D)$ generation rate, and (b1)-(b4) the $O(^1D)$ loss rate as a function of the $Ar/O_2$ gas mixture ratio (90/10, 80/20, 50/50, 10/90). The discharge conditions are the same as those in figure 15.

Figure 14 illustrates the changes of the fluxes of the three neutral particles $Ar^*$, $O(^3P)$, and $O(^1D)$ on the driven electrode (first column) and the grounded electrode (second column) with varying gas mixtures for $N = 2$ at pressures of 200 mTorr, 300 mTorr, and 500 mTorr. With the increasing pressure, the background gas ($O_2$ and Ar) density increases, leading to an enhancement of the flux of $O(^3P)$. However, as the pressure increases, the fluxes of $O(^1D)$ and $Ar^*$ decrease. From our previous work [23], this is because the loss rate for $O(^1D)$ and $Ar^*$ increase significantly, which leads to a decrease of the $O(^1D)$ and $Ar^*$ densities and fluxes.

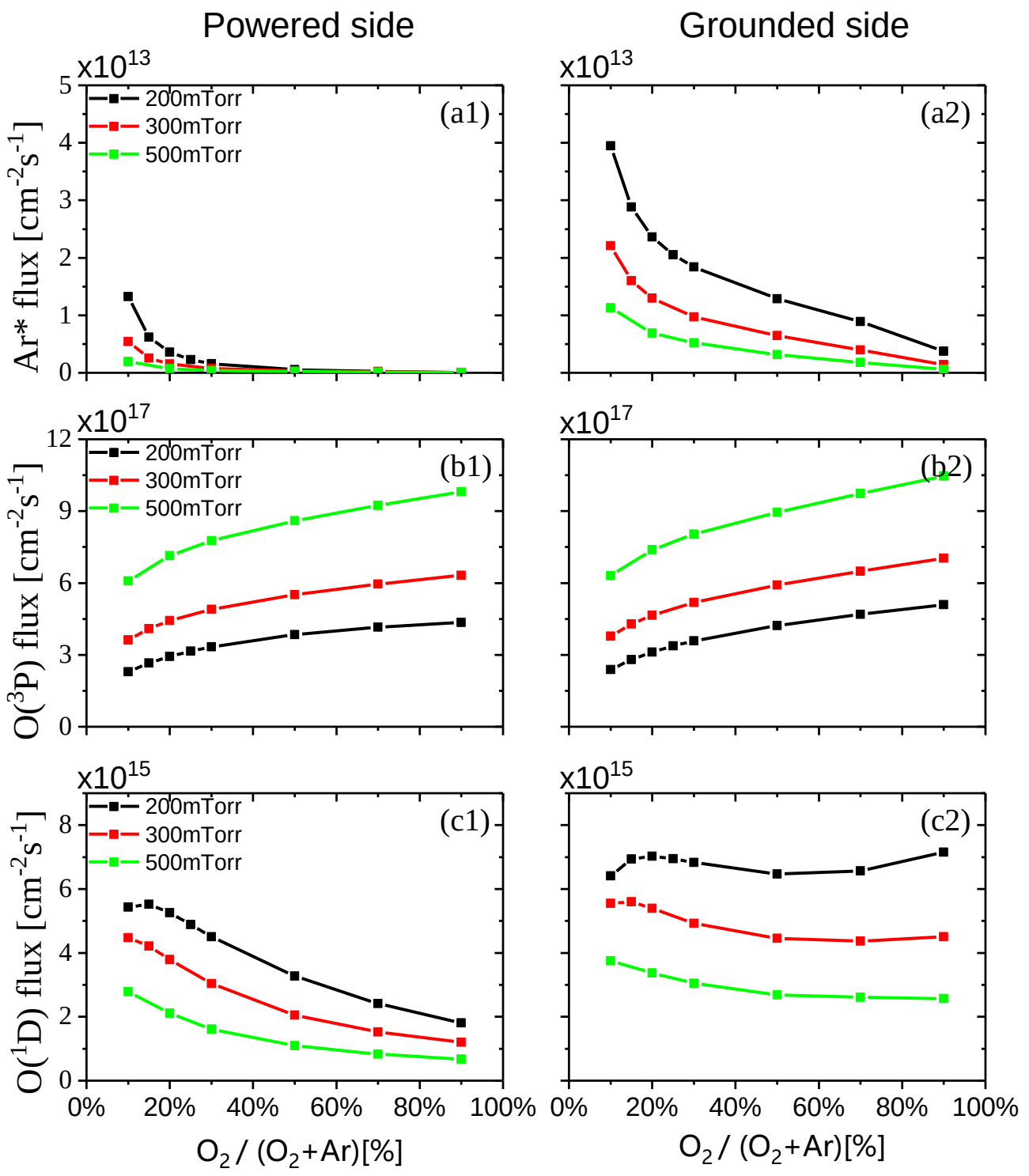

**Figure 14** Fundamental frequency period averaged fluxes of (a1)-(a2) $Ar^*$, (b1-(b2) $O(^3P)$ and (c1)-(c2) $O(^1D)$, at 200 mTorr (black line), 300 mTorr (red line), 500 mTorr (green line) near the powered electrode and grounded electrode sustained by a dual frequency ($N$ = 2) voltage waveform. Discharge conditions: Ar/$O_2$ (90/10~10/90) with an electrode gap of 3 cm and a fundamental frequency of 13.56 MHz with 150 V voltage amplitude.

From an industrial perspective, using tailored voltage waveform provides significant additional control of neutral fluxes to the electrodes in Ar/$O_2$ gas mixtures. By adjusting the waveform shape through changing the number of consecutive harmonics it is synthesized from, a spatial asymmetry is induced. This spatial asymmetry has distinct effects on the fluxes of neutrals at both electrodes. These effects are markedly different for different species. In this way the fluxes of one neutral particle species to the wafer can be enhanced, while the flux of another species is kept constant or is reduced. Using sawtooth up voltage waveforms and placing the substrate on the grounded electrode and using a higher number of harmonic frequencies allows realizing higher fluxes of deposition precursors ($O(^1D)$). And it is also found that the discharge at an $O_2$ admixture of about 20% can generate almost the same flux of $O(^1D)$ as that produced for an $O_2$ admixture of 90%, which implies that the industry can utilize this phenomenon to enhance the $O(^1D)$ flux to the electrodes at low $O_2$ admixture.

## 4 Conclusions

This study focuses on electron dynamics, and the transport of charged particles and neutrals in the presence of different numbers of harmonic frequencies *N* used to synthesize sawtooth up tailored driving voltage waveform, at different pressures and Ar/$O_2$ gas mixtures in capacitively coupled plasmas. As the number of harmonic frequencies increases, the electron density not only increases significantly at the sheath boundary but also in the bulk region due to the increase of the electron power absorption rate, leading to a decrease in electronegativity. Ionization mostly happens at one distinct time within the fundamental RF period (0.1 *T*) in the bulk region and near the grounded electrode. Meanwhile, the density of electrons changes from a spatially symmetric distribution to an asymmetric distribution, with low energy electrons present near the powered electrode and high energy electrons located near the grounded electrode. Moreover, the amplitude of the DC self-bias voltage generated via the EAE decreases as a function of *N*, because the rapid sheath expansion at the powered electrode provides sufficient energy to electrons, which cause an ionization peak near the powered electrode, which induces a plasma asymmetry, as there is no such enhanced ionization peak at the grounded electrode.

As the pressure increases, the electronegativity decreases. The peak magnitude of the electric field in the bulk region does not change significantly, but its spatial position varies. As the pressure increases, there are peaks in ionization both in the bulk region and near the grounded electrode, and the peak in the discharge center is almost equal to the peak near the grounded electrode. This is related to the electron heating dynamics and the decreased mean free path of electrons with increasing pressure. The absolute value of the DC self-bias decreases at higher pressure, primarily because more ions are generated near the bulk region and the powered electrode rather than only near the powered electrode.

As the $O_2$ admixture increases, the electronegativity is enhanced. The number of high-energy electrons significantly increases, and their distribution gradually shifts towards the grounded electrode. Ionization mainly occurs around 0.1 *T* at different gas ratios. For a low $O_2$ admixture, ionization is evident near the sheath expansion phase, in the bulk region, and near the sheath collapse phase at the grounded electrode, exhibiting a hybrid α and DA mode. As the $O_2$ admixture increases, the ionization during the sheath expansion phase at the powered electrode and in the bulk region are attenuated and the discharge gradually changes to a DA-dominated mode. Moreover, the amplitude of the DC self-bias voltage increases due to a change of the spatial distribution of the ionization rate.

Under the influence of a slope asymmetric driving voltage waveform, the flux of oxygen-related neutrals to both electrodes can be adjusted by VWT. This modulation is particularly pronounced under conditions of low oxygen content and weak electronegativity. An increase of the electronegative oxygen admixture diminishes this effect. This is primarily due to the higher electronegativity causing the discharge to localize, resulting in significant differences in generation rate, and loss rate compared to scenarios with lower electronegativity. Regarding argon-related neutrals, there is a notable disparity between the two electrodes due to the electrical asymmetry. This asymmetry causes the peak of the generation rate to shift towards the grounded electrode. Consequently, the neutral flux decreases significantly on the powered electrode side compared to single-frequency discharges, while it increases on the grounded electrode. From an industrial standpoint, employing sawtooth up voltage waveforms and positioning the substrate near the grounded electrode while using a higher number of harmonic frequencies can enhance the deposition of precursor such as $O(^1D)$. A discharge with an oxygen admixture of approximately 20% can yield nearly equivalent levels of $O(^1D)$ flux as a discharge with a 90% oxygen admixture. This finding suggests that process engineers can effectively increase the $O(^1D)$ flux without increasing the $O_2$ admixture.

In this paper, the discharge characteristics and the transport of charged particles and neutrals have been simulated and analysed under different discharge conditions, which might provide a reference for the industry. In the future, the effects of $Ar/O_2$ gas mixture discharge driven by tailored voltage waveform on the deposition of thin films will be studied to optimize the deposition morphology.

## Acknowledgments

This work was supported by the National Natural Science Foundation of China under Grant Nos. 12020101005, 11975067, 12347131, the Liaoning Provincial Natural Science Foundation Joint Fund under Grant Nos. 2023-BSBA-089, the China Scholarship Council (No. 202306060179), and the German Research Foundation via the SFB 1316 (project A4) as well as grant 428942393.

## Data availability statement

All data that support the findings of this study are included within the article (and any supplementary files).

## ORCID iDs

Wan Dong https://orcid.org/0000-0002-0724-1697
Li Wang https://orcid.org/0000-0002-3106-2779
Chong-Biao Tian https://orcid.org/0009-0003-5902-8278
Yong-Xin Liu https://orcid.org/0000-0002-6506-7148
Yuan-Hong Song https://orcid.org/0000-0001-5712-9241
Julian Schulze https://orcid.org/0000-0001-7929-5734)